\documentclass[a4paper,12pt]{article}
\pdfoutput=1
\usepackage{graphicx, rotating}
\usepackage{hyperref}\usepackage{slashed}
\usepackage{ifpdf}

\ifx\pdfoutput\undefined
\usepackage[dvips,bookmarks]{hyperref}	
\else
\usepackage{hyperref}	
\fi
\hypersetup{colorlinks,bookmarksopen,bookmarksnumbered,citecolor=verdes,
linkcolor=blus,pdfstartview=FitH,urlcolor=rossos}
\def\hhref#1{\href{http://arxiv.org/abs/#1}{#1}} 

\newcommand{\cm}{\,{\rm cm}}

\usepackage{multicol}
\usepackage{color}
\definecolor{rosso}{cmyk}{0,1,1,0.4}
\definecolor{rossos}{cmyk}{0,1,1,0.55}
\definecolor{rossoc}{cmyk}{0,1,1,0.2}
\definecolor{blu}{cmyk}{1,1,0,0.3}
\definecolor{blus}{cmyk}{1,1,0,0.6}
\definecolor{bluc}{cmyk}{1,1,0,0.1}
\definecolor{verde}{cmyk}{0.92,0,0.59,0.25}
\definecolor{verdec}{cmyk}{0.92,0,0.59,0.15}
\definecolor{verdes}{cmyk}{0.92,0,0.59,0.4}

\font\tenrsfs=rsfs10 at 12pt
\font\sevenrsfs=rsfs7
\font\fiversfs=rsfs5
\newfam\rsfsfam
\textfont\rsfsfam=\tenrsfs
\scriptfont\rsfsfam=\sevenrsfs
\scriptscriptfont\rsfsfam=\fiversfs
\def\mathscr#1{{\fam\rsfsfam\relax#1}}

\newcommand{\fig}[1]{~\ref{fig:#1}}
\newcommand{\eq}[1]{~{\rm (\ref{eq:#1})}}

\newcommand{\GeV}{\,{\rm GeV}}
\newcommand{\TeV}{\,{\rm TeV}}
\newcommand{\eV}{\,{\rm eV}}

\def\circa#1{\,\raise.3ex\hbox{$#1$\kern-.75em\lower1ex\hbox{$\sim$}}\,}

\newcommand{\eflux}{$e^++e^-$}

\newcommand{\eqref}[1]{(\ref{#1})}

\newcommand{\NP}{Nucl. Phys.}

\newcommand{\beq}{\begin{equation}}
\newcommand{\eeq}{\end{equation}}
\newcommand{\MeV}{\,{\rm MeV}}
\def\circa#1{\,\raise.3ex\hbox{$#1$\kern-.75em\lower1ex\hbox{$\sim$}}\,}
\makeatletter

\def\art{\@ifnextchar[{\eart}{\oart}}
\def\eart[#1]#2#3#4#5#6{{\rm #2}, {#3 #4} {\rm (#6) #5} [arXiv:\-{\hhref{#1}}]}
\def\hepart[#1]#2{{\rm #2, arXiv:\-\hhref{#1}}}
\newcommand{\oart}[5]{{\rm #1}, {#2 #3} {\rm (#5) #4}}

\newcounter{alphaequation}[equation]
\def\thealphaequation{\theequation\hbox to
0.6em{\hfil\alph{alphaequation}\hfil}}
\def\eqnsystem#1{
\def\@eqnnum{{\rm (\thealphaequation)}}
\def\@@eqncr{\let\@tempa\relax \ifcase\@eqcnt \def\@tempa{& & &} \or
\def\@tempa{& &}\or \def\@tempa{&}\fi\@tempa
\if@eqnsw\@eqnnum\refstepcounter{alphaequation}\fi
\global\@eqnswtrue\global\@eqcnt=0\cr}
\refstepcounter{equation} \let\@currentlabel\theequation \def\@tempb{#1}
\ifx\@tempb\empty\else\label{#1}\fi
\refstepcounter{alphaequation}
\let\@currentlabel\thealphaequation
\global\@eqnswtrue\global\@eqcnt=0 \tabskip\@centering\let\\=\@eqncr
$$\halign to \displaywidth\bgroup \@eqnsel\hskip\@centering
$\displaystyle\tabskip\z@{##}$&\global\@eqcnt\@ne
\hskip2\arraycolsep\hfil${##}$\hfil& \global\@eqcnt\tw@\hskip2\arraycolsep
$\displaystyle\tabskip\z@{##}$\hfil
\tabskip\@centering&\llap{##}\tabskip\z@\cr}
\def\endeqnsystem{\@@eqncr\egroup$$\global\@ignoretrue} \makeatother

\begin{document}

\begin{center}
IFUP-TH/2009-9\hfill 

\bigskip\bigskip\bigskip

{\huge\bf\color{magenta}
Dark Matter Interpretations\\
of  the  $e^\pm$ Excesses after FERMI}\\

\medskip
\bigskip\color{black}\vspace{0.6cm}
{
{\large\bf Patrick Meade$^a$, Michele Papucci$^a$, \\[3mm] Alessandro Strumia$^b$ and Tomer Volansky$^a$}
}
\\[7mm]
{\it $^a$ Institute for Advanced Study, Princeton, NJ 08540} \\
{\it $^b$ Dipartimento di Fisica dell'Universit{\`a} di Pisa and INFN, Italia} \\

\bigskip\bigskip\bigskip\bigskip

{
\centerline{\large\bf Abstract}

\begin{quote}
The cosmic-ray excess observed by PAMELA in the positron fraction and by FERMI and HESS in $e^-+e^+$ can be interpreted in terms of DM annihilations or decays into  leptonic final states. 
Final states into $\tau$'s or $4\mu$ give the best fit to the excess.
However, in the annihilation scenario, they are incompatible with photon and neutrino constraints, 
unless DM has a quasi-constant density profile.
Final states involving $e$'s are less constrained but poorly fit the excess,
unless hidden sector radiation makes their energy spectrum smoother,
allowing a fit to all the data with a combination of leptonic modes.
In general, DM lighter than about a TeV cannot fit the excesses, so PAMELA should {fi}nd a greater positron fraction 
at higher energies. The DM interpretation can be tested by FERMI $\gamma$ observations above 10 GeV:
if the $e^\pm$ excess is everywhere in the DM halo, inverse Compton scattering on ambient light produces a well-predicted $\gamma$ excess that FERMI should soon detect.
\end{quote}}

\end{center}

\newpage

\tableofcontents

\section{Introduction}
Recently the PAMELA experiment observed an unexpected rise with energy of the $e^+/(e^++e^-)$ fraction in cosmic rays, suggesting the existence of a new component.
The sharp rise suggests that the new component may be visible in the  $e^- + e^+$ spectrum.
As  worked out in \cite{Moiseev:2007js} and further stressed in~\cite{CKRS}, FERMI  provides the first precise measurement of the $e^++e^-$ spectrum.
The main purpose of this article is to analyze the implications of the first FERMI $e^++e^-$ data~\cite{FERMI}, in conjunction with the new HESS measurements~\cite{HESSepm}.

Although the peak hinted by previous data~\cite{ATIC-2} is not confirmed,
the FERMI and HESS~\cite{HESSepm} observations still demonstrate a deviation from the naive power-law spectrum, indicating an
anomalous excess compared to conventional background predictions of cosmic ray fluxes at the Earth.
Still, it is important to note that predictions of cosmic ray fluxes are model dependent and fraught with large uncertainties, therefore gauging the exact significance or the correct interpretation is quite difficult.
Astrophysical explanations have been put forth in an attempt to explain the data~\cite{pulsars}:
the most promising candidates are pulsars, which produce an excess with an energy spectrum that ranges from smooth to peaked to rastered.
Thereby the FERMI measurement of the energy spectrum does not clarify the nature of the excess, and the
most interesting possibility remains open: the excess could be the first manifestation of Dark Matter, rather than
a new background to Dark Matter searches.

\smallskip

In this paper we concentrate on the Dark Matter (DM) interpretation.  As we show below, the PAMELA, FERMI and HESS observations still point towards non-conventional DM models, with
leptophilic final states, large ``event rates" in our galaxy and a high mass scale ($\mathcal{O}$(TeV)).  
The 
reason for this is traced back to the fact that the new FERMI data does not show any sharp feature at low scale.  Such a feature must be visible if the DM mass is low.  Therefore, {\it DM lighter than about a TeV is now excluded as an interpretation of the PAMELA excess}.  This conclusion further implies that the positron fraction should plateau or continue rising, but cannot go down at the higher energies being probed by PAMELA.
Thus the DM scenario may be excluded in the future by PAMELA.

Many models with these properties have been put forth to explain the excesses,
and the range of possibilities span various annihilating and decaying dark matter scenarios.
DM annihilations need to be enhanced with respect to what naively is suggested by thermal freeze-out in standard cosmology.    
This can be achieved either through  the Sommerfeld effect \cite{Somm,CKRS,Nima}, some resonance \cite{CKRS,murayama}, 
DM sub-clumps \cite{Kuhlen:2008aw}, or, of course, by a non-standard cosmology.
It is therefore important to not only distinguish between the  DM scenario and other astrophysical ones, but also to differentiate between the various DM models themselves.

Indeed, while many models can explain the PAMELA anomaly, a large fraction of the existing models are inconsistent with other astrophysical measurements.  Previous studies~\cite{CKRS, BCST, mpv, berkeley, Bergstrom:2008ag} have analyzed some of the constraints and here we revisit the analysis once more in light of the new results.    In particular, we consider the following set of measurements,
\begin{itemize}
\item PAMELA positron fraction~\cite{PAMELA}, anti-proton fraction~\cite{PAMELApbar}, and preliminary electron flux~\cite{waseda}.
\item FERMI~\cite{FERMI} and HESS~\cite{HESSepm} $e^++e^-$ flux.
\item FERMI preliminary diffuse $\gamma$ rays~\cite{FERMIgamma}.
\item HESS and VERITAS observations of $\gamma$ rays from the Galactic Center~\cite{HessGC} (including preliminary new HESS data), the Galactic Ridge~\cite{HessGR}
and dwarf galaxies~\cite{HessSgrDwarf,VERITAS}.
\item {\sc SuperKamiokande} up-going neutrino-induced $\mu^\pm$ flux~\cite{SK, SKshower}.
\end{itemize}

\medskip

\begin{figure}
\begin{center}
$$\hspace{-0.02\textwidth}\includegraphics[width=1.05\textwidth]{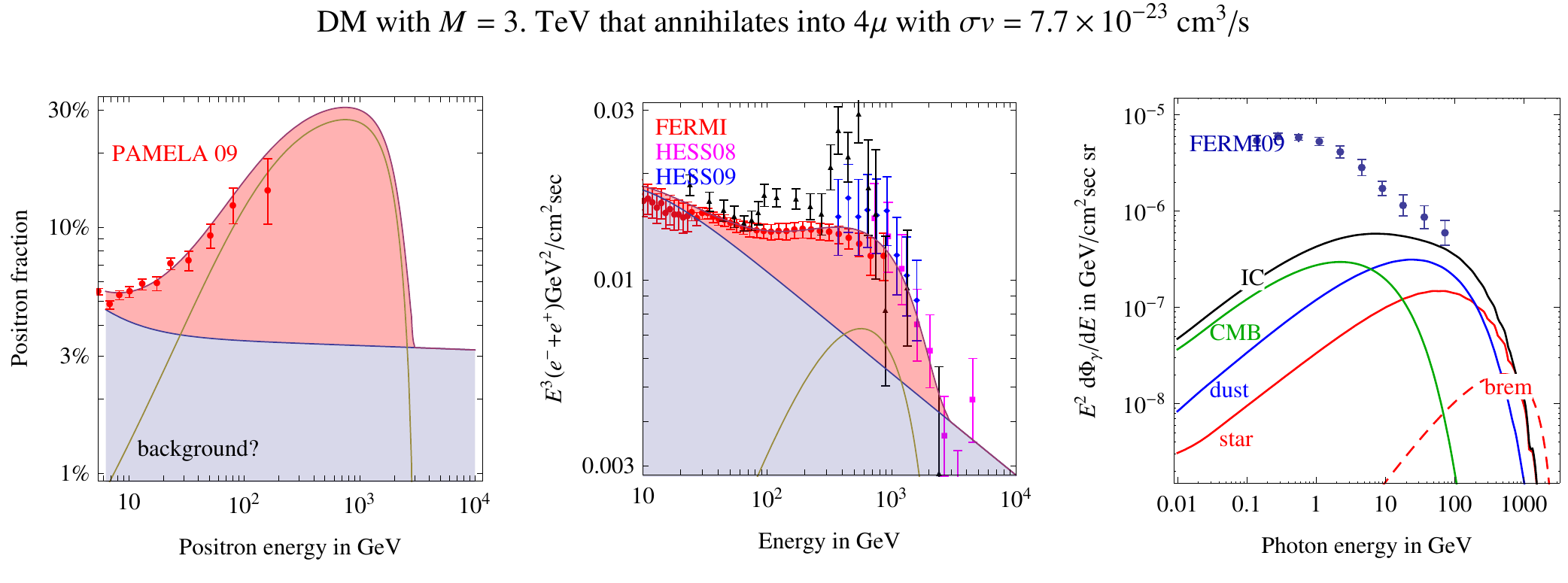}$$
$$\hspace{-0.02\textwidth}\includegraphics[width=1.05\textwidth]{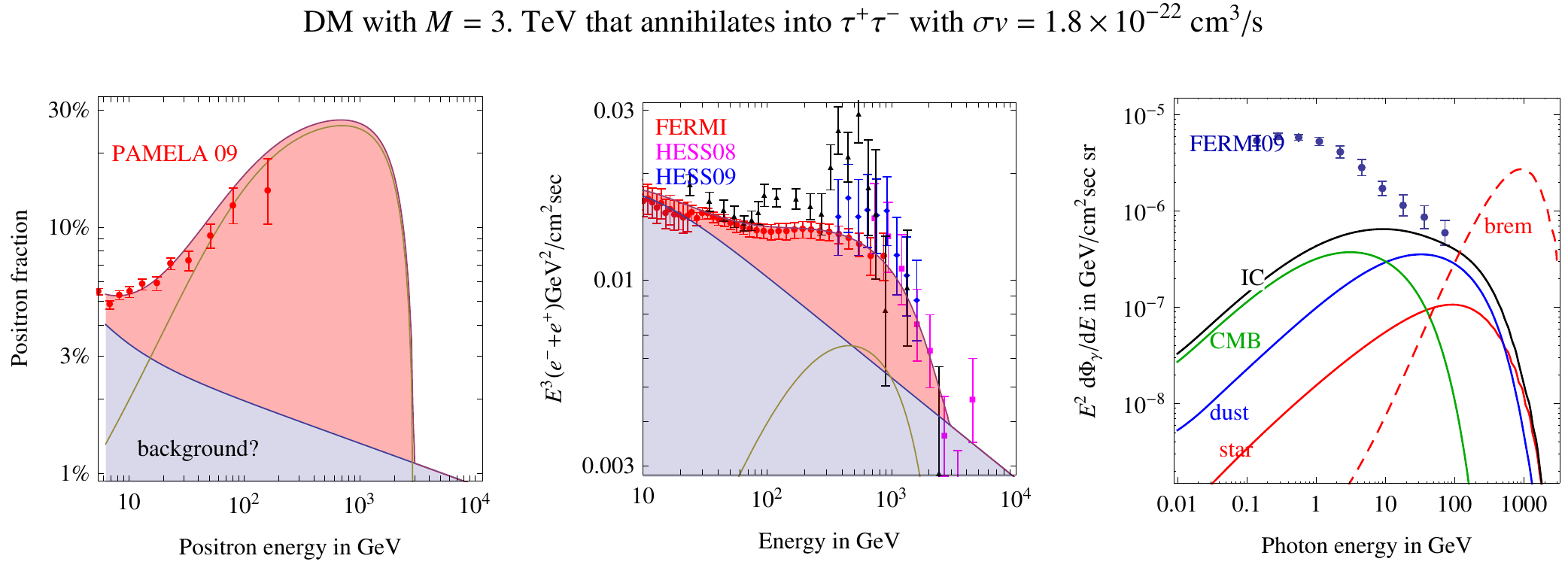}$$
\vspace{-0.5cm}
\caption{\em {\bf Sample DM fits}.
In the upper (lower) row we consider DM annihilations into $\mu^+\mu^-\mu^+\mu^-$ ($\tau^+\tau^-$)
with  MED diffusion~\cite{minmedmax} and the NFW (isothermal) DM profile:
all good fits are very similar.
Left: the positron fraction compared with the PAMELA excess.
Middle: the $e^++e^-$ flux compared with the FERMI and HESS data.
Right: the DM contribution to the diffuse photon energy spectra produced by bremsstrahlung (dashed red curve) 
and IC (black thick line); we also separately show the 3 IC components from star-light (red),
CMB (green), dust (blue).
\label{fig:sample}}
\end{center}
\end{figure}

The measurements above are insufficient to verify the DM picture.  Ideally what one would hope for is some positive signal in a direct detection experiment.  Nevertheless, there may still be important indirect evidence for DM by future measurements involving photons.  These channels may shed light on whether the DM story is fully self consistent or not.  In particular FERMI will measure additional diffuse gamma ray fluxes in our galaxy.  This is an important measurement for two reasons.  
First, one expects DM to typically clump into subhalos within our Galaxy (as shown by $N$-body simulations and theoretical arguments).  
These subhalos in turn can be a source of DM annihilation into channels with associated gamma rays and would provide an unambiguous signature for DM~\cite{Kuhlen:2008aw}.  
Second, barring the model dependency of including subhalos, the general DM profile extends above the galactic disk and should produce photon signals from Final State Radiation
(FSR), Inverse Compton Scattering
(ICS) and synchrotron.  If DM exists in the form suggested by the charged particle fluxes, the signal from ICS in the diffuse gamma ray background as measured by FERMI should be an unambiguous clue as to whether DM is the cause of the excesses.

\medskip

In this paper we study DM that is consistent with all data including the preliminary PAMELA electron measurement, 
most recent FERMI  electron plus positron  measurements and the preliminary diffuse $\gamma$ measurements.  We then make predictions for the additional FERMI diffuse $\gamma$ ray measurements and the future PAMELA positron fraction.  
In fig.\fig{sample} we demonstrate a sample good fit of the PAMELA and FERMI data and its prediction for the ICS spectrum that is an outcome of our analysis.  The ICS flux in this example is representative of most models that fit PAMELA, FERMI and HESS.

This paper is organized as follows.  In Section~\ref{sec:modspace} we describe the various final states in our DM calculations.  We relate these final states to various models of DM including annihilating, decaying, and long lived intermediate particles.   In Section~\ref{sec:ics} we outline the calculation for the contribution to the diffuse gamma ray background from ICS and synchrotron.  We point out certain  relations between the calculation of ICS and diffusion and bring up certain issues not discussed previously in the literature.  We discuss a semi-analytic approximation to Inverse Compton scattering that illuminates some basic physical features.  The reader might want to jump directly to Section~\ref{sec:results}, where we fit the various models to the existing cosmic ray data,
 identifying  the parameter space viable for explaining PAMELA after the recent FERMI and HESS results.  
In view of the strong restrictions we make precise predictions for the upcoming diffuse gamma ray measurement by FERMI and comment on whether this will be a sufficient measurement to test the DM explanation.  In the appendices, we describe some of the details of our approach to calculating neutrino fluxes and fitting procedure.

\section{Dark Matter models}\label{sec:modspace}

In this section we outline the ``model" space that we cover in our study.  Despite the existence of a cornucopia of models~\cite{Nima,decaying,susy,UEDsucks,modsusyandothers,annihilatelow,Hall}, we attempt to be as complete as possible in our scan while taking into account known constraints~\cite{CKRS,BCST,mpv,berkeley}.  By and large, DM models bifurcate into annihilating or decaying DM.  In order to further explain the significant electronic activity and absence of hadronic activity, one either assumes a symmetry or otherwise the production of light states which are kinematically forbidden to decay into hadrons.  In the latter case, the number of SM final states must be four or more.  From the constraints point of view, the number of final states dictates the hardness of the injection spectrum leading to the following breakdown of the space of models:

\begin{itemize}
\item Annihilating dark matter
\begin{itemize}
\item Final states with 2 SM particles
\item Final states with $\geq$ 4 SM particles
\end{itemize}
\item Decaying dark matter
\begin{itemize}
\item Final states with 2 SM particles
\item Final states with $\geq$ 4 SM particles
\end{itemize}
\end{itemize}

The bifurcation into annihilating versus decaying DM has consequences for indirect detection because fluxes depend on the DM density $\rho$, as $\rho^2$ and $\rho$ for annihilating and decaying models respectively.  This implies that annihilating  DM models are more constrained by the non-observation of a $\gamma$ excess from the Galactic Center, where the DM density is large  (this bound can be weakened by assuming that DM annihilates into a long-lived particle~{\cite{Rothstein:2009pm}).
Decaying DM models~\cite{decaying} have a lot more freedom.  The $e^\pm$ excesses can be accommodated by choosing the DM decay rate, which unlike the thermal DM annihilation rate, is not linked to cosmology.
Unfortunately, there is no definitive signal that allows one to differentiate between the decaying and annihilating DM scenarios.   With this in mind our scan of model space will focus more on annihilating DM, but nevertheless include bounds and predictions for decaying DM.   The relation between the rates of the two follows from a straightforward substitution: $\rho^2 \langle \sigma v\rangle/2M^2 \leftrightarrow \rho \Gamma/M$, where $\tau=1/\Gamma$ is the DM lifetime.  

For annihilating DM, final states with 2 SM particles such as $W^+W^-$ or $e^+e^-$ are found in more conventional models of WIMPS~\cite{susy,UEDsucks}.  These models were already highly constrained prior to the release of the new FERMI data~\cite{CKRS} and additionally require a non-thermal cosmology or large boost factor.  They typically lead to either sharp features in the $e^++e^-$ flux or large antiproton fluxes.  The large antiproton fluxes are incompatible with the PAMELA data unless the scale is pushed to~$\mathcal{O}(10)$ TeV or maybe, with extremal astrophysical parameters,  below $\sim 200$ GeV~\cite{CKRS}.   The sharp features in the \eflux flux that were a benefit in fitting the ATIC~\cite{ATIC-2} data are now disfavored as we show in Section~\ref{sec:results}.  

\smallskip

The $\geq$ 4 SM particle final state is a very interesting possibility for explaining the recent excesses.  As has been noted, one needs a leptophilic final state {\em and} a large event rate.  This can be obtained assuming that DM interacts with a lighter particle with mass $m_\phi$, such that a
long range attractive force generates a Sommerfeld enhancement to DM annihilations at low velocities.
This lighter particle can no longer be the $W$ boson~\cite{CKRS}, as DM annihilations into $W^+W^-$ does not provide a viable
interpretation of the HESS and FERMI data.  
One possibility is to introduce an ad hoc new particle, for example by gauging $L_\mu - L_\tau$~\cite{CKRS}, allowing to explain the leptonic activity.
A more elegant model of Sommerfeld-enhanced leptophilic DM annihilations is obtained~\cite{Nima} by assuming
that the new particle only interacts with DM and is lighter than the proton,
so that kinematics allow the new particles to decay only into the lighter leptons and pions.

There are two leading effective operators of how a light hidden gauge group could couple to the SM and thus allow annihilations into visible SM particles.  These are
\begin{equation}\label{kinetic}\label{higgs}
\epsilon F_{\rm DM}^{\mu\nu}F_{\mu\nu}\qquad \hbox{and}\qquad \epsilon h^2  h_{\rm DM}^2
\end{equation}
where $F_{\rm DM}^{\mu\nu}$ is the field strength of the dark gauge group, 
 $h$ is the SM Higgs, and $h_{\rm DM}$ is a Higgs of the dark gauge group.  
The first operator describes the kinetic mixing of electric charge in our sector with the hidden kinetic term, and the second operator
represents mass mixing through the Higgs. While both terms allow for DM annihilations into the light dark state, 
the subsequent mixing of the DM light state with SM particles determines whether it couples through charge (kinetic mixing with the photon)
 or mass (mass mixing with the higgs).  
 In either case $m_\phi \leq1\,\mathrm{GeV}$ is needed to  explain the leptophilic final states, but the branching fractions are different, as they are proportional to the square of either the electric charge or the final state mass respectively.

DM annihilations into two light hidden sector particles
 produce at least 4 body final states of SM particles.
 Indeed, as pointed out in~\cite{mpv}, if 
 the light hidden sector particle is charged under a hidden gauge group that is nonabelian (as preferred for phenomenological reasons~\cite{Nima}), 
final state radiation in the hidden sector can produce more hidden sector particles which subsequently decay into $> 4$ SM final states {\em and} can significantly effect final energy distributions.   
Alternatively, if the hidden sector has several higgs or multiple light particles there could be a cascade decay producing $\geq 4$ SM final states.  Cascade decays and showering will not give identical spectra~\cite{mpv,berkeley}, but since there is a great deal of model dependency in cascade decays we will only analyze the case of FSR.  
This can be computed in terms of the unknown gauge coupling constant in the DM sector, $\alpha_{\rm DM}$,
and we consider several values for it 
to demonstrate the effects of multiple final light states that eventually decay to the SM.

To demonstrate those models that are compatible with FERMI and PAMELA we choose to simulate the final states of $2\mu$, $2\tau$, $4e$, $4\mu$,  $4\pi^\pm$,  $4\tau$, and additionally shower to create final states with $>4$ SM particles.  Other final states have been considered in the past \cite{CKRS, mpv} and were shown to be excluded.  In section~\ref{sec:results} we concentrate on the bounds for each channel assuming these final states as fully exclusive, but we will also provide the constraints for models where the branching fractions are given assuming couplings proportional to charge.  Other models which can generate $\geq 4$ SM final states but with different BFs then those explicitly shown in our paper~\cite{annihilatelow} have bounds that can be inferred from a linear combination of our exclusive channels.

If the light particle can decay into $\tau^+\tau^-$, it can also decay into $p\bar p$.
However due to kinematical reasons, neglecting final state radiation, 
all protons have energy $E\ge m_p  M/m_\phi$, which, for sufficiently high DM mass, $M$, is large enough to push the  signal to energies
not yet probed by PAMELA~\cite{mpv}.
A hidden sector with light particles in the few-GeV mass range that decay into multi-$\tau$ modes is presently motivated by the CDF multi-muon anomaly~\cite{CDFmu}.  
Since light particles cannot be much heavier than $2m_\tau$ (we fix $m_\phi=4\GeV$), the $\tau$ spectra are not broad, and consequently the
resulting $e^\pm,\gamma,\nu$ spectra are just slightly broader than in the corresponding $2\tau$ cases with DM mass rescaled by 1/2.  
However, $\tau$ gives a large $\gamma$ flux (due to the $\pi^0$ decay chain), 
so that the $\tau$ cases are {\bf{\em only}} compatible with bounds on  astrophysical $\gamma$ fluxes~\cite{mpv} if DM decays
(rather than annihilates) or if DM has a quasi-constant density profile (disfavored by $N$-body simulations).

\section{Testing the $e^\pm$ Excess with Diffuse Gamma Rays}\label{sec:ics}
So far observations indicated excesses in cosmic ray $e^\pm$.
Unfortunately, the directionality of $e^\pm$ is lost by the galactic magnetic fields making $e^\pm$ 
almost isotropic in the sky.  Observations of  $e^\pm$ excesses therefore hardly allow one to test if it is present only locally or everywhere in the Milky Way.  
To have an unambiguous determination of the source direction we would need to observe a related excess in neutrinos or photons.

Electrons generated by DM lose essentially all their energy via Inverse Compton scattering, $e^\pm\gamma \to e^{\prime\pm}\gamma'$, on ambient light with average energy $E_\gamma \sim \eV$.  Such scatterings give rise to photons with larger energy $E_{\gamma'} \sim E_\gamma(E_e/m_e)^2  \sim10\GeV$, which is in the energy range being probed by FERMI.  As discussed below, this DM  ICS $\gamma$ flux is only marginally affected by astrophysical and DM distribution uncertainties.  The  reasons for this can be traced back to two observations: (i)  Far away from the Galactic Center, the DM uncertainties are relatively mild.  (ii)  As we will see in Section~\ref{sec:results}, all DM models that fit the data predict roughly the same $e^\pm$ spectrum, as it is now mostly fixed by the new measurements (given the new FERMI and HESS results).  Thereby the DM ICS spectrum is well predicted.   As already illustrated in fig.\fig{sample}
it is not much below the first FERMI diffuse $\gamma$-ray data, released for energies $\leq 10$ GeV in a specific angular region. Therefore, if the $e^\pm$ excess is due to DM, FERMI is expected to observe an associated $\gamma$ excess which is not sensitive to the specific DM model or DM density profile.  Whether such an excess is seen or not, will decisively implicate on the DM (or any other mechanisms that produces $e^\pm$ in a spherical region away from the galactic plane) interpretation of the measured excesses. Alternative scenarios involve $e^\pm$ generated locally (e.g.\ by a powerful pulsar) or
along the galactic plane (e.g.\ by supernov\ae).

The ICS signal was computed in previous works in relation to the PAMELA measurement
by running the GALPROP code~\cite{galprop} with some given astrophysical parameters~\cite{Neal}.
Here we perform our own computation discussed in the rest of this section, where we
justify our statement that the DM ICS signal has minor astrophysical uncertainties.
We bring up some details not discussed previously in the literature on this subject.

\subsection{$e^\pm$ Energy Losses}
\medskip

Energy losses of $e^\pm$ in the galaxy are due to two competing processes:
\begin{itemize}
\item[1)]  Synchrotron radiation in the galactic magnetic fields $B$
(with energy density $u_B =B^2/2$ in natural units).
\item[2)] 
IC scatterings on ambient photons, with energy density $u_\gamma$.
\end{itemize}
The energy loss rate is given by,
\beq\label{eq:b(E)}
 - \frac{dE_e}{dt} =b(r,z,E)= \frac{4\sigma_T }{3 m_e^2}  
 E_e^2\left[ u_B + \sum_{i=1}^3 u_{\gamma i}  \cdot R_i(E_e)\right]  \eeq
where $\sigma_T = 8\pi  r_e^2/3$ with $r_e = \alpha_{\rm em}/m_e$.
\begin{figure}[t]
\begin{center}
\includegraphics[width=0.45\textwidth]{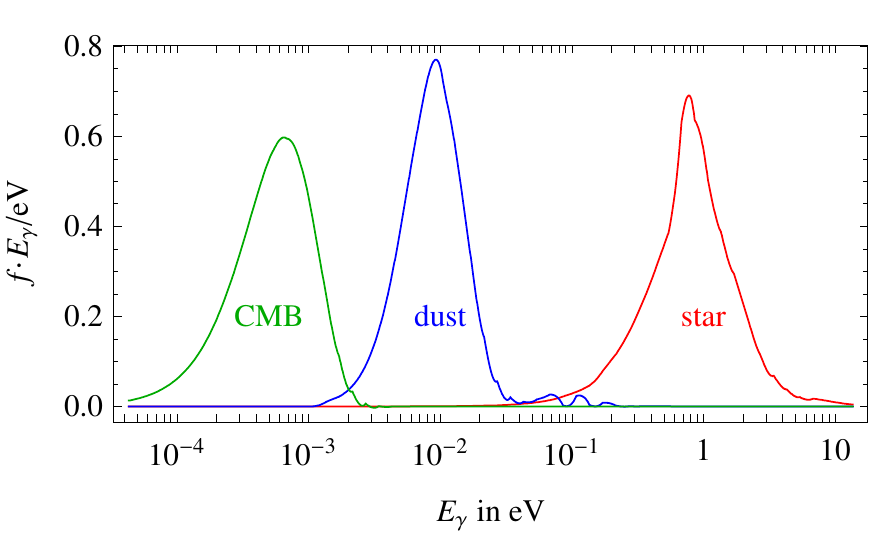}\qquad
\includegraphics[width=0.45\textwidth]{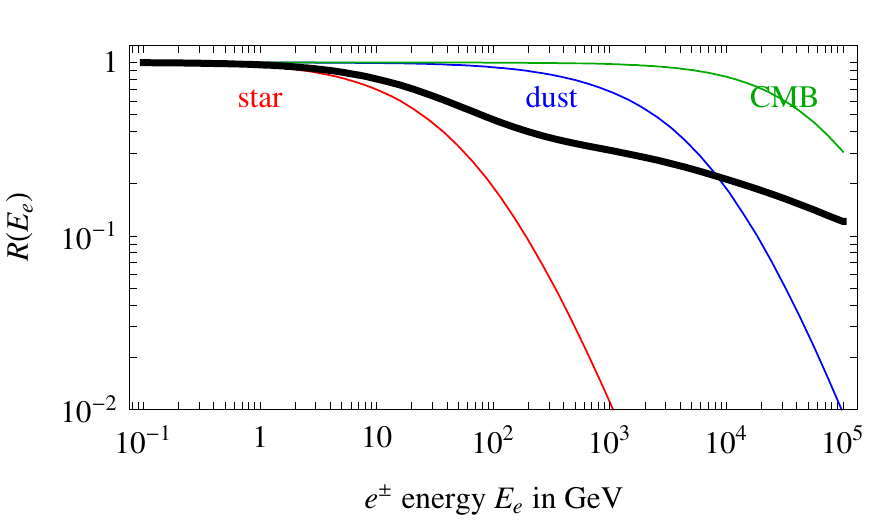}
\caption{\em Left: Energy spectra of the 3 galactic light components~\cite{isrf}, normalized to unity.
Right: The functions $R_i(E_e)$ which encode the relativistic corrections to the ICS energy loss.  The black think line shows the function $R(E_e)$ defined below Eq.~(\ref{eq: JG}).
\label{fig:fR}}
\end{center}
\end{figure}
 The light profiles $u_\gamma(r,z)$ are relatively well known, being composed of three components
(hence the sum over $i$ in the above equation):
\begin{enumerate}
\item A constant CMB with energy spectrum
$du/dE_\gamma = E_\gamma^3/\pi^2/(e^{E_\gamma/T}-1)$ and mean energy $\langle E_{\gamma\rm CMB}\rangle=0.6$ meV.

\item Star-light concentrated in the galactic center, with optical mean energy $\langle E_{\gamma \rm star}\rangle=1.1\eV$.

\item Star-light re-scattered by dust, with mean energy $\langle E_{\gamma\rm dust}\rangle=0.01\eV$.
\end{enumerate}
We plot the spectral shape of the three components in Fig.\ref{fig:fR}a.
The functions $R_i(E_e)$ in the above eq.\eq{b(E)} encode the relativistic correction to the non-relativistic Thompson limit
of IC scattering
and are plotted in Fig.\fig{fR}b.
They equal to unity at $E_e \ll  m_e^2/\langle E_{i\gamma}\rangle$ and exhibit a $ E_e^{-2}$
suppression at higher energies.
In practice, this relativistic effect must be taken into account for ICS on star-light for $e^\pm$ energies, $E_{e}\circa{>} 250\GeV$.

\medskip

Magnetic fields, on the other hand,  are not precisely known: they likely lie between $1$ and $10\mu{\rm G}$,
and may have the approximated profile~\cite{galprop},
\begin{equation}
\label{eq: Brz}
B(r,z) \approx 11\mu{\rm G} \cdot\exp(-r/10\,{\rm kpc} - |z|/2\,{\rm kpc}).
 \end{equation}
They give rise to energy losses into synchrotron radiation probed down to microwave frequencies.
Thereby, while we know that essentially all the $e^\pm$ energy goes into photons,
it is hard to establish the precise relative proportion of the two effects discussed above.
Assuming the form of $B(r,z)$ given in Eq.~(\ref{eq: Brz}), Fig.\fig{u} demonstrates the energy densities $u_B$ and $u_{\gamma i}$.  We see that synchrotron energy losses are everywhere subdominant by about one order of magnitude.
In view of the relativistic effect discussed in the
previous paragraph there is only one possible exception:
synchrotron and ICS give comparable energy losses in the inner few kpc at $E_e \circa{>}\TeV$,
where ICS energy losses on star-light are suppressed by the relativitic factor $R_{\rm star}$.

\smallskip

\subsection{Computing $\gamma$ from Inverse Compton}

The IC process $e^\pm\gamma\to e^{\pm\prime}\gamma'$  scattering of an $e^\pm$ with energy $E_e$ and isotropic initial direction on isotropic $\gamma$ with energy $E_\gamma$ gives  $\gamma'$ with the following $\gamma'$ energy spectrum~\cite{ICth}:
\begin{eqnarray}
 \label{eq:dNIC}
\frac{dN'_\gamma}{d E'_\gamma~dt}& =&2\pi r_e^2 \frac{m_e^2 }{E_e^2} \frac{u_\gamma}{E^2_\gamma}f_{\rm IC}(q,\epsilon),
\\
f_{\rm IC}(q,\epsilon )&=&2q\ln q+(1+2q)(1-q)+\frac{1}{2}\frac{(\epsilon q)^2}{1+\epsilon q}(1-q)  .
\end{eqnarray}
Here  $E_e\gg m_e$ and $\epsilon$, $\Gamma$, $q$ are the dimensionless variables defined as:
\beq \epsilon=\frac{E'_\gamma}{E_e},\qquad \Gamma=\frac{4E_\gamma E_e}{m_e^2},\qquad q= \frac{\epsilon}{\Gamma(1-\epsilon)}.\eeq
$E_{\gamma'}$ lies in the range $E_\gamma/E_e \le \epsilon\le \Gamma/(1+\Gamma)$.
The non-relativistic (Thompson) limit corresponds to $\Gamma\ll1$, so that $\epsilon\ll 1$, the last term in $f_{\rm IC}$ is negligible,
and $0\le q\le 1$.
The total energy loss rate in eq.\eq{b(E)} is recovered
by integrating eq.\eq{dNIC} over $E'_\gamma$.

\begin{figure}[t]
\begin{center}
\includegraphics[width=\textwidth]{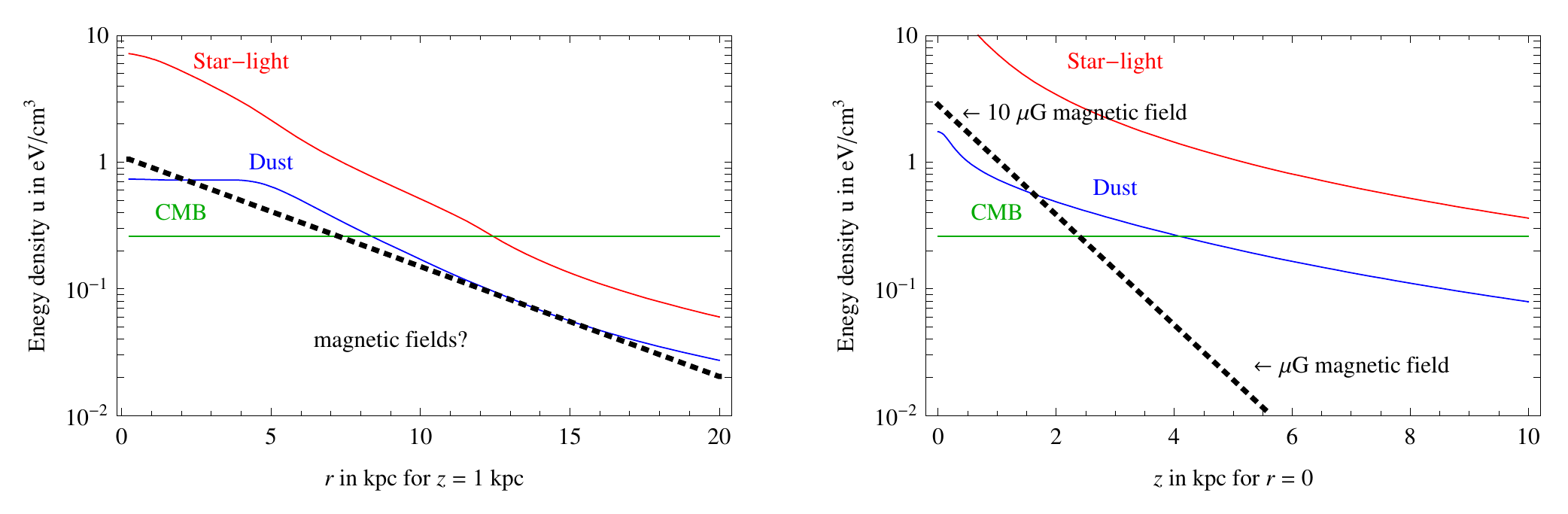}
\caption{\em Slices of the energy density profiles $u(r,z)$ of 
star-light (red upper curve) re-scattered by dust (blue), CMB (green horizontal line),
and of the presumed galactic magnetic fields (dashed). Left: Profiles as a function of the radius $r$, for a fixed $z=1$ kpc.  Right: Profiles as a function of the height, $z$, for a fixed $r=0$. The values corresponding to $B=1$ and $10\mu{\rm G}$ are indicated in the right plot.
\label{fig:u}}
\end{center}
\end{figure}

\medskip
The ICS flux $\Phi_\gamma = dN_\gamma/dS\,dt$ obtained for a given line of sight and generic energy and spatial distributions of initial $e^\pm$ and $\gamma$, is found to be,
\beq 
\frac{d\Phi_{\gamma'}}{dE_{\gamma'} d\Omega} = \frac{1}{2}\alpha_{\rm em}^2
\int_{\rm l.o.s.} ds \int\!\! \int \frac{dn_e}{dE_e}\frac{du_\gamma}{dE_\gamma}
 \frac{dE_e}{E_e^2} \frac{dE_\gamma}{E_\gamma^2}f_{\rm IC}.
 \label{eq:Compton1}
\eeq
Here we do not include the anisotropic correction to~\ref{eq:Compton1} due to the fact that light is emitted preferentially from the galactic plane.  The effect in the FERMI region is $\mathcal{O}(10\%)$~\cite{anisotropic}.
The DM ICS flux depends on the $e^\pm$ density $f=dn_e/dE$.
It is computed by numerically solving the diffusion-loss equation 
\beq \label{eq:diffeq}-
K(E)\cdot \nabla^2f - \frac{\partial}{\partial E}\left[ b(r,z,E) f \right] =
\frac{1}{2} \left(\frac{\rho}{M}\right)^2 \langle \sigma v\rangle \frac{dN_{e}}{dE}
\eeq
where $dN_e/dE$ is the spectrum produced by one DM annihilation while $b(r,z,E)$ is given in Eq.~\eqref{eq:b(E)}.  For more details on the above diffusion-loss equation and its approximate solutions see e.g.~\cite{FornengoDec2007, MDM3,mpv} and references therein.
On the practical level, we work on a grid in cylindrical coordinates $r,z,E$  and discretize the differential equation\eq{diffeq} and its boundary conditions into a system of thousands of linear equations, easily solved with Mathematica.  We take into account the spatial dependence of the energy loss term $b$ assuming that each of the light components $u_{\gamma i}$, $u_B$, can be factorized in space and energy as discussed below.

\subsection{Approximating the Diffuse $\gamma$-ray Flux}

In order to gain more intuition, it is beneficial to study the ICS spectrum in an a approximate manner.   Let us, therefore, obtain a simplified expression for the diffuse gamma-ray flux which illuminates the physics. 
The photon spectrum is the sum of 3 components (star-light, dust, CMB)
which have position-dependent intensity, and roughly the same energy spectra at any point:
$dn_\gamma/dE_\gamma = \sum_i  f_{\gamma i}(E_\gamma) u_{\gamma i}(r) /\langle E_{\gamma i}\rangle$
where $u_{\gamma i}$ is the total energy density, and the functions $f_{\gamma i}$ are plotted in fig.\fig{fR}a and
normalized to unity: $\int dE_{\gamma } f_{\gamma i} = 1$.

Assuming that $e^\pm$ diffusion is negligible\footnote{In the halo-function formalism~\cite{FornengoDec2007,MDM3} this corresponds to assuming 
unity halo-fuction, $I\simeq 1$.}
the energy spectrum of $e^\pm$ generated by DM has the same shape at any position:
\beq 
\frac{dn_{e^-}}{dE}=\frac{dn_{e^+}}{dE}=  \frac{3m_e^2}{4\sigma_T u_{\rm tot}}
 \frac{N_e(E)}{E^2}\times \frac{\sigma v}{2} \left(\frac{\rho(R)}{M}\right)^2,\qquad
 N_e(E)=
 \int_{E}^{M}  dE'~\frac{dN_{e}}{dE'}  \cdot  I,
\label{eq:fluxpositrons}
\eeq
where $I$ is defined as in~\cite{FornengoDec2007,MDM3}.
Therefore the ICS spectrum simplifies to
\beq \label{eq:ICapprox}
\frac{d\Phi_{\gamma'}}{dE_{\gamma'}} =\sum_i G_{i\rm IC}(E'_\gamma) J_{i\rm IC} \frac{9   r_\odot \langle\sigma v\rangle}{64\pi \langle E_{\gamma i}\rangle}
 \left(\frac{ \rho_\odot }{M}\right)^2\eeq
where the dimensionless factors  $J_{\rm IC}$ and $G_{\rm IC}$ respectively encode astrophysics and particle-physics:
\beq
\label{eq: JG}
J_{i\rm IC}=\int d\Omega
 \int_{\rm l.o.s.} \frac{ds}{r_\odot} \left(\frac{\rho(r)}{\rho_\odot}\right)^2 \frac{u_{\gamma i}}{u_{\rm tot}},\qquad
 G_{i\rm IC}= m_e^4  \int\!\! \!\!\int N_e(E_e)f_{\gamma i}(E_\gamma)
 \frac{dE_e}{E_e^4}  \frac{dE_\gamma}{E_\gamma}\frac{ f_{\rm IC}}{R(E_e)}   .
\eeq
We explicitly see that the overall amount of energy $u_{\rm tot} =  u_B + \sum_i u_{\gamma i}$
does not matter, but only the ratios among the different components, up to the relativistic correction
$R(E_e) \equiv  [u_B + \sum_i u_{\gamma i} R_i(E_e)]/u_{\rm tot}$. 
Unfortunately, in general this last factor is position-dependent and therefore $G_{iIC}$ is
not purely determined by particle-physics.
Nevertheless, this latter dependence is weak and to leading order $G_{iIC}$ encodes the particle physics information which enters the diffuse spectrum.
In fact, assuming that IC scattering on star-light dominates $e^\pm$ energy losses
up to an energy $E_e^*$, it is given by
$R(E_e) \sim  R_{\rm star} (\min(E_e,E_e^*))$,
where astrophysics is all condensed in the parameter $E_e^*$,
which presumably is large enough not to be crucial.
The function $R$ is plotted in fig.\fig{fR}b for the FERMI region.

\medskip

It is interesting to note that in the case where synchrotron energy losses are subdominant with respect to the ICS energy losses, $\sum_i J_{i\rm IC} = J$, where $J$ is the usual factor that encodes the astrophysics of 
DM annihilations into photons (with energy spectrum $dN_\gamma/dE_\gamma$ per DM annihilation):
\beq 
\label{gammaflux}
\frac{d \Phi_\gamma}{dE_\gamma} = J\frac{r_\odot \langle \sigma v\rangle}{8\pi}
 \frac{dN_\gamma}{dE_\gamma} 
 \left(\frac{ \rho_\odot }{M}\right)^2 ,\qquad
J = \int d\Omega \int_{\rm l.o.s.} \frac{ds}{r_\odot} \left(\frac{\rho(r)}{\rho_\odot}\right)^2 .
\eeq
In table~\ref{tab:J} we list the astrophysical $J$ factors for the `$10^\circ\div20^\circ$'  region observed by FERMI.  It is indeed apparent that $J$ amounts to the bulk part of the contribution.  Moreover, since the region probed is far from the GC, the various $J_i$ factors only weakly depend on the DM profile and the predicted spectrum is therefore robust.

\begin{table}
$$\begin{array}{c|ccc|ccc}
\hbox{region} & \multicolumn{3}{c|}{\hbox{DM annihilation}} & \multicolumn{3}{c}{\hbox{DM decay}}  \\ 
10^\circ<|b|<20^\circ  & \hbox{NFW} & \hbox{Einasto} & \hbox{isothermal}
 & \hbox{NFW} & \hbox{Einasto} & \hbox{isothermal}\\ \hline
 \hbox{$J_{\rm star,IC}$} & 4.6 & 6.0 & 2.9 & 2.4 & 2.6 & 2.2\\
 \hbox{$J_{\rm dust,IC}$} &1.1 & 1.4 & 0.78 & 0.67 &0.70 & 0.63\\
 \hbox{$J_{\rm CMB,IC}$}&1.0& 1.2& 0.86   & 1.4 & 1.4 & 1.5\\ \hline
 \hbox{$J$} & 7.4 & 9.2 & 5.0 & 4.9 & 5.0 & 4.7
\end{array}$$
\caption{\em {\bf Astrophysical factors} $J_i$ for Inverse Compton within the `$10^\circ\div20^\circ$' region observed by FERMI.
Since the region is large and away from the GC, the $J_i$ only have a minor dependence on the DM profile.\label{tab:J}}
\end{table}%

As a word of caution we stress that the $J$ integral should be computed only inside the diffusion volume.
For the above table we assumed the MED propagation parameters ($L=4\,{\rm kpc}$), and, in the case of the `$10^\circ\div20^\circ$' region
this cut  reduces $J$ by about $10\%$ with respect to its full volume value ($L=\infty$).
A significant reduction is found only for the implausible MIN configuration  ($L=1\,{\rm kpc}$).\footnote{
We compute ICS neglecting the contribution of $e^\pm$ generated outside the diffusion cylinder.
Indeed, despite energy losses due to ICS, their mean free path is $\approx100$ kpc for $E\approx 1\TeV$ and $u \approx \eV/\cm^3$,  larger enough than the size of our galaxy, that their 
contributions to the ICS $\gamma$ flux gets mostly lost.}
Finally, throughout our analysis below, we used both the approximation discussed above and the exact solution to the diffussion-loss equation.  We found that the approximation for the ICS fluxes differs by  a factor of ${\cal O}(2)$.  This is illustrated in Fig.~\ref{fig:icscomp}.

\begin{figure}
\begin{center}
\includegraphics[width=0.45\textwidth]{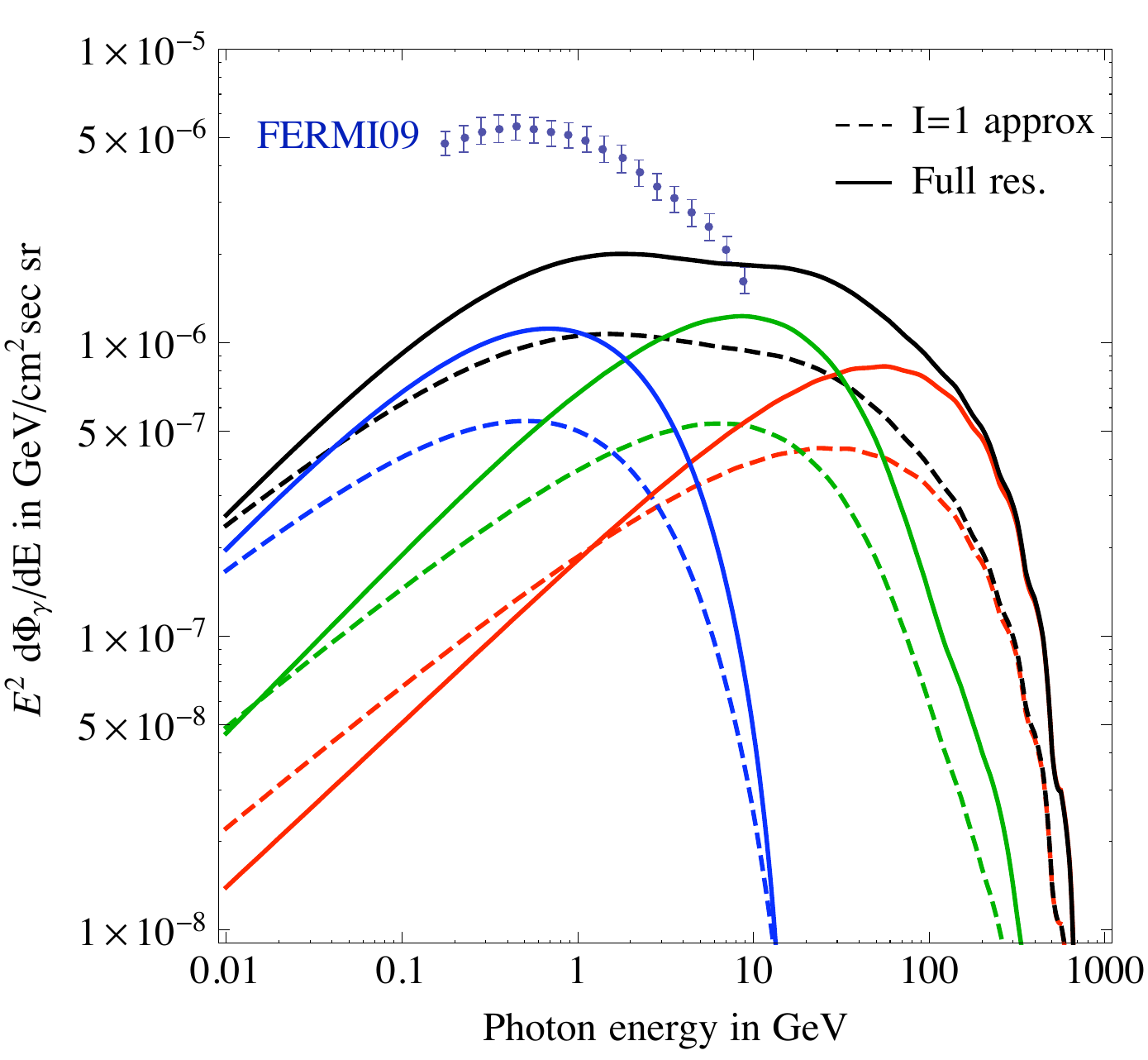}
\caption{\em {\bf Inverse Compton: exact vs approximated.} We compare our full calculation for ICS (solid curves)
with the diffusion-less
$I=1$ approximation of eq.\eq{ICapprox} (dashed curves)
for the `$10^\circ\div 20^\circ$' region and DM annihilating into $\mu^+\mu^-$ with $\sigma v=10^{-22} cm^3/s$ and NFW with MED propagation.
Additionally we plot the individual contributions to ICS from CMB (blue), dust (green) and starlight (red), while the total contribution is in black.  
The ICS approximation is good within a factor of two.
\label{fig:icscomp}}
\end{center}
\end{figure}

   The plots in Fig.\fig{sample}c show the result of our full ICS computation.
We have separately  shown the DM predictions
for the three ICS components within the `$10^\circ\div20^\circ$' region so far observed by FERMI.  The DM model plotted gives a best fit to the FERMI, PAMELA and HESS $e^\pm$ excess.
In fact, as mentioned above, the electron spectra turn out to be rather similar for the majority of models that fit FERMI, PAMELA and HESS and therefore the ICS spectra are very similar in the various cases.  We provide more evidence to this statement in section~\ref{sec:results}.
This observation is in contrast to the FSR $\gamma$ fluxes (red dashed lines in Fig.\fig{sample}) which are highly model dependent \cite{BCST,mpv,berkeley} and typically  dominate over ICS at higher energies close to the DM mass.

\medskip

In conclusion, the prediction for the ICS $\gamma$ flux is robust: in the `$10\div 20$' region
the DM contribution should be visible at $E_\gamma\circa{>}100\GeV$, and other regions offer better sensitivities.  Reducing the IC
DM signal would require unexpected astrophysics:
either very large galactic magnetic fields (such that the detectable signal anyhow moves from ICS to synchrotron radiation) or perhaps a very thin diffusion cylinder.

\section{Interpreting the PAMELA and FERMI Observations}\label{sec:results}
We now demonstrate the viable models that can fit all the available data for charged cosmic rays {\em including} the new FERMI and HESS data. For our analysis we fit to the PAMELA $e^+/(e^+ + e^-)$ data and the  $e^++e^-$ data from FERMI and HESS.  
We also include the preliminary un-normalized PAMELA $e^-$ spectrum  available at~\cite{waseda}, although it has a minor impact.   We then describe the predictions and bounds for ICS as discussed in Section~\ref{sec:ics}, and include bounds from several other observations previously studied in~\cite{BCST,HisanoNu, MDM3,mpv}.  These include the bounds from the HESS photon measurements in the Galactic Center and Galactic Ridge and from up-going muons measured at SuperKamiokande.   

 {\em The FERMI data is conservatively fitted} adding in quadrature statistical with systematic uncertainties
independently for each data-point.
We consider uncertainties on the smooth DM halo profile $\rho(r)$, $e^\pm$ propagation,
and the spectral index and normalization of the $e^+$ and of the $e^-$ astrophysical backgrounds.
We keep fixed the local DM density, $\rho_\odot=0.3\GeV/\cm^3$.
Changing it would be equivalent to an overall rescaling of the DM annihilation or decay rate, which renormalizes in the same way all indirect DM observables.  Therefore, the comparison between the regions favored by the $e^\pm$ excesses and the constraints from $\gamma$ and $\nu$
observations remains fully meaningful.
 In Appendix~\ref{fit} we describe further how our fit is performed, and here we simply summarize the main points.

Other, less established bounds, are not included.  In particular, 
ref.~\cite{CiPa} finds that angular regions distinct from the one observed by FERMI and so far observed only by EGRET, provide stronger constraints.  
We do not use here the controversial EGRET observations.
Once FERMI will present data corresponding to other regions, it will be easy to establish bounds with the use of our approximation described in the previous section.  This is done by simply rescaling our predictions for the `$10^\circ\div20^\circ$'  region using the new $J$ factors for the additional (yet unknown) regions.
Furthermore, we do not consider the `WMAP haze'~\cite{WMAPHaze} which is a hint that a possible excess in synchrotron radiation could be due to DM.  The haze has been shown to be consistent with a wide variety of DM masses and final states and therefore will not constrain the space of models compared to other measurements.  It would however, be interesting to study in more detail the precise predictions for the haze for those models that can fit the rest of the data.

\medskip

\begin{figure}
\begin{center}
\includegraphics[width=0.45\textwidth]{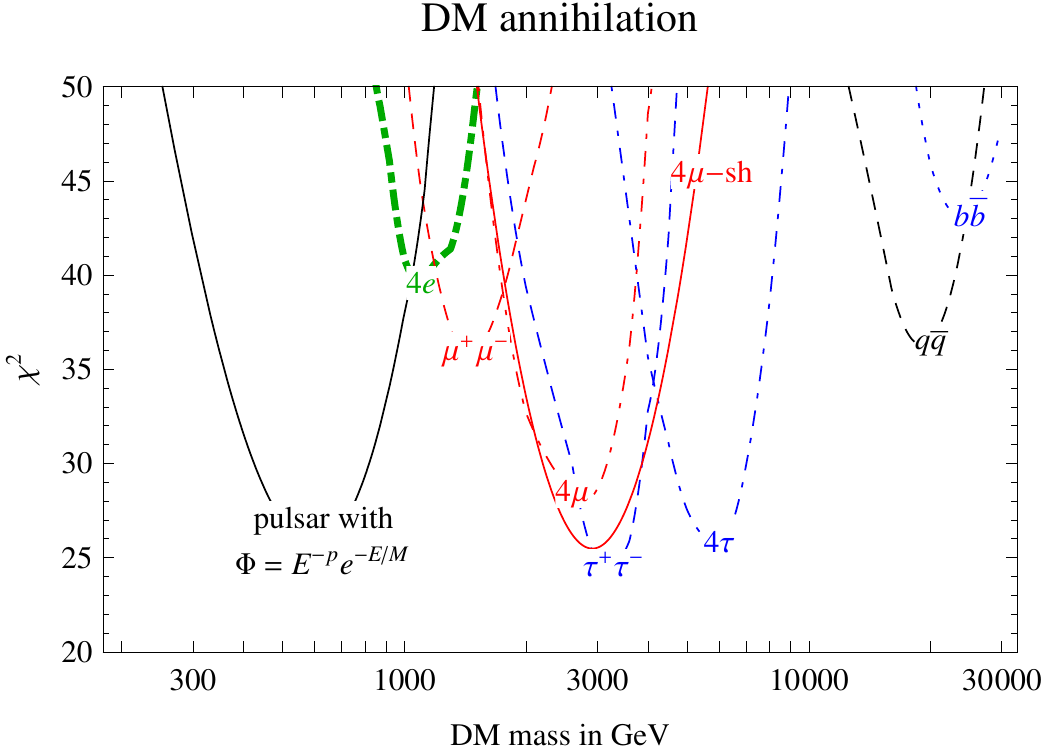}\hspace{0.05\textwidth}
\includegraphics[width=0.45\textwidth]{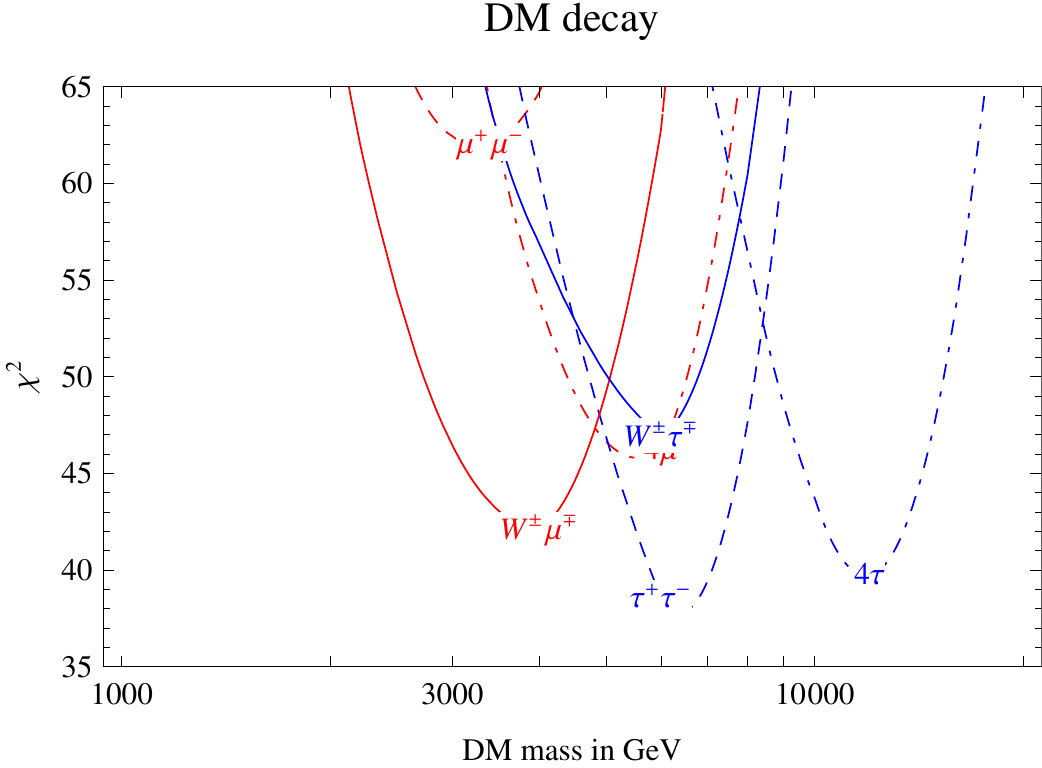}
\caption{\em {\bf Global fit} to PAMELA, FERMI and HESS data.
The labels on each curve indicate the primary DM annihilation (left) or decay (right) channel.  In the left panel a hypothetical flux from a pulsar is also plotted, with an assumption that the flux is given by $\Phi=E^{-p} e^{-E/M}$.  In the left panel all final states for DM annihilation do not include hidden sector FSR, except for the curve labelled $4\mu-sh$.  This curve demonstrates that by including the hidden sector shower the $\chi^2$ is significantly improved and is as good of fit as any other hypothesis.
\label{fig:fit}}
\end{center}
\end{figure}

Fig.\fig{fit} shows the $\chi^2$ as function of the DM mass for various DM annihilation (left) or DM decay (right) modes.
We find that, independently of the non-observation of an excess in the $\bar p$ PAMELA data,
only some leptonic modes can reproduce all data.  Here HESS observations play a key role demanding that the $e^\pm$ excess terminates in a sharper way than what typical of non-leptonic channels, irrespectively of the DM density profile.
DM heavier than 10 TeV that annihilates or decays into light quarks still provides a reasonable fit to PAMELA and FERMI data,
if  fitted conservatively.  However it is disfavored by the HESS $e^++e^-$ data, and presumably the photon data as well ( in the annihilating case).

The only spectral feature that can allow one to discriminate the various modes lies at the high end of  the spectrum between 1 and 3 TeV, where we only have the electron HESS data which is less precise than the FERMI data.
The leptonic channels can be ordered according to the sharpness of their end-point:
\beq 2e ~>~ 4e\sim 2\mu ~>~ 4\mu \sim 2\tau ~\circa{>}~ 4\tau .\eeq
The main difference with respect to the previous ATIC $e^++e^-$ data~\cite{ATIC-2} is that a peak is no longer present, and therefore the $2e$ mode (namely, DM annihilations or decays into $e^+ e^-$) is now excluded independently of the photon bounds, since it predicts a too sharp end-point.  All other leptonic modes we consider provide comparably good fits to the data, that therefore cannot discriminate which (combination of) modes is the correct one.
We also explored the spectra produced by polarized $\mu$ or $\tau$, similarly finding that present data
do not allow to discriminate the various possibilities.

\medskip

The FERMI data alone, becomes more constraining if one takes into account the appropriate correlations among the systematic uncertainties at different data points.
Assuming that systematics only affects the overall energy scale, and that our description of the background is still valid
at the level of precision of FERMI statistical errors, data would imply a preference for the smoother $4\mu$ or $2\tau$ modes.
As stated above, assuming instead that FERMI data must be fitted conservatively (so that they are consistent with a power law with no spectral features)
and dropping the HESS data (which implies a termination of the excess at $\sim2\TeV$), non-leptonic modes
are consistent with the data if DM is heavier than $\sim10\TeV$ and provide fits as good as the leptonic modes
for $M\sim 30\TeV$.
In particular, Minimal Dark Matter (which predicted the PAMELA $e^+$ excess as
Sommerfeld-enhanced DM annihilations into $W^+W^-$
with $M\approx 9.6\TeV$ and no ATIC peak~\cite{MDM3}) is no longer a viable DM interpretation of the  $e^++e^-$ excess, 
if HESS or FERMI  conclusively establish that the excess terminates around 2 TeV (or if the DM profile is found to be sufficiently dense at the GC to be constrained by HESS).

\smallskip

Therefore a future more precise measurement of the $e^++e^-$ spectrum around the end-point of the excess at $\sim2\TeV$
will be very important to disentangle the DM annihilation modes.
This is illustrated by the best-fits shown in  fig.\fig{ICsamples}b for the $4\ell$ annihilation modes.
We here fixed MED propagation and the Einasto DM density profile, so that each best-fit could be further slightly improved by adjusting the propagation and the DM profile. 
We see that the various channels mildly differ at the $(2\div 3)\TeV$ energy at which the $e^++e^-$ excess terminates. 
We also see that in all cases the other feature present in the FERMI data, namely
the mild hardening of the $e^++e^-$ spectral slope around 100 GeV, can be simply attributed
to the takeover of the DM component, without indicating any special spectral feature in it.
Finally, all best fits predict that the the positron fraction must continue to go up to about 1 TeV in an essentially unique way dictated by the FERMI data.

\medskip

\subsection{Best $e^\pm$ Fits vs $\gamma$ and $\nu$ Bounds}
Figures\fig{boundse}, \fig{boundsee} and \fig{boundsDecay} show the best-fit regions 
in the ($M,\langle\sigma v\rangle$) plane for DM annihilations and in the $(M,\tau)$ plane for DM decays.
Each panel assumes a specific mode and a specific DM profile.
In each panel the red regions are favored by the global fit of FERMI, HESS and PAMELA data at $3$ and $5\sigma$ (2 dof).
The green bands are favored by PAMELA (at $3\sigma$ for 1 effective dof;
we here used the lowest-energy FERMI bins to fix the overall flux, finding 
a $\langle\sigma v\rangle$ which is a few times lower than the best-fit bands of~\cite{BCST}.  This could only be due to distinct theoretical computations of the $e^\pm$ fluxes and the fit to the ATIC peak).

The plots show that the best-fit DM mass and annihilation cross section or decay rate
is very different in the various cases, ranging from 1 TeV (for the $\mu^+\mu^-$ channel)
to almost 10 TeV (for the $4\tau$ channel). 
Thereby the channel could be discriminated looking at different observables.
In view of the multi-TeV DM masses, it will be difficult to test these DM scenarios at LHC.

The most promising observables are the  $\gamma$ and $\nu$ fluxes generated by DM annihilations or decays.
As all data for the moment shows no excess, we compare the regions that best fit the $e^\pm$ excesses
with the regions excluded by other observations.
Even if the $e^\pm$ best-fit regions are almost the same irrespectively of the DM density profile,
such comparisons must be done for a given DM density profiles.  Indeed, photon measurements point to their source, and so the associated bounds strongly depend on the DM density profile.
We consider the NFW~\cite{NFW}, Einasto~\cite{Einasto} and isothermal~\cite{isothermal} DM density profiles.

\begin{figure}[p]
\begin{center}
$$\includegraphics[width=0.99\textwidth]{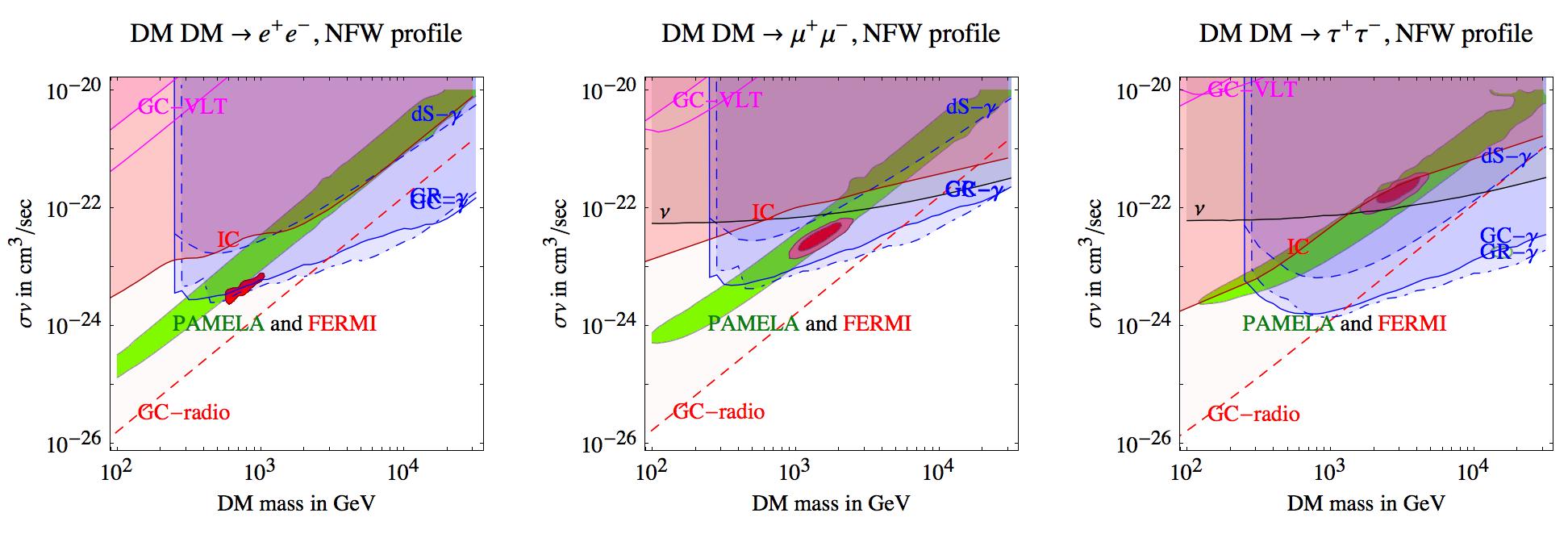}$$
$$\includegraphics[width=0.99\textwidth]{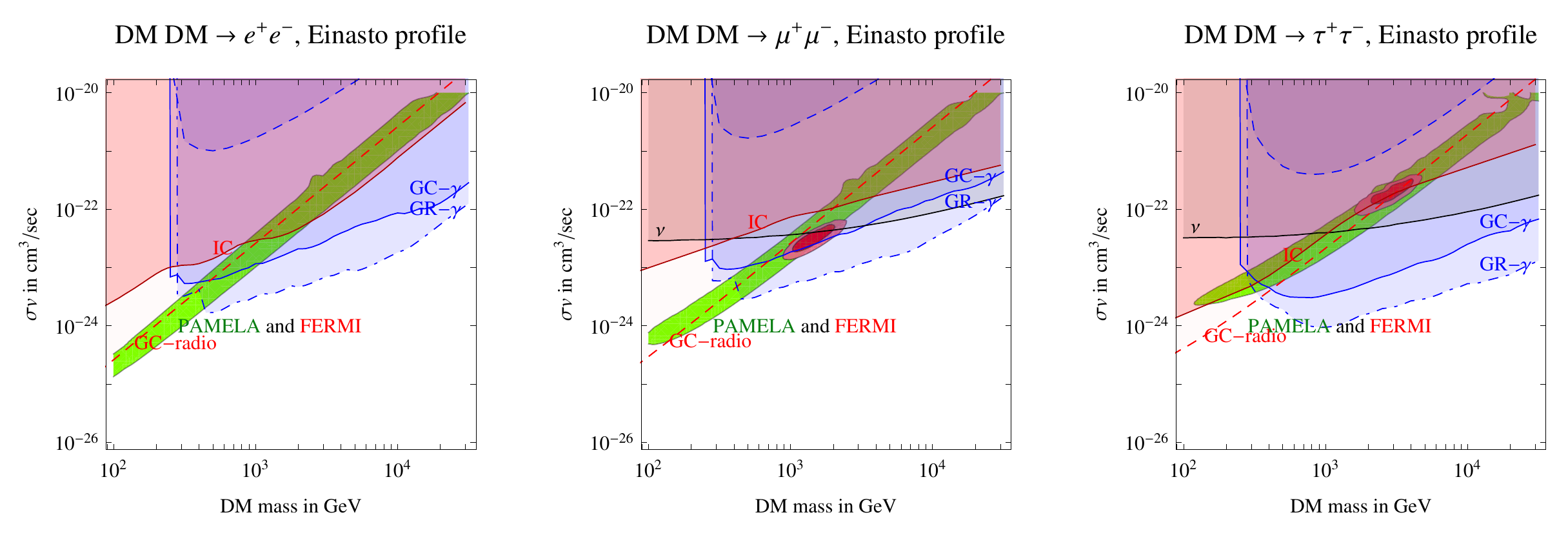}$$
$$\includegraphics[width=0.99\textwidth]{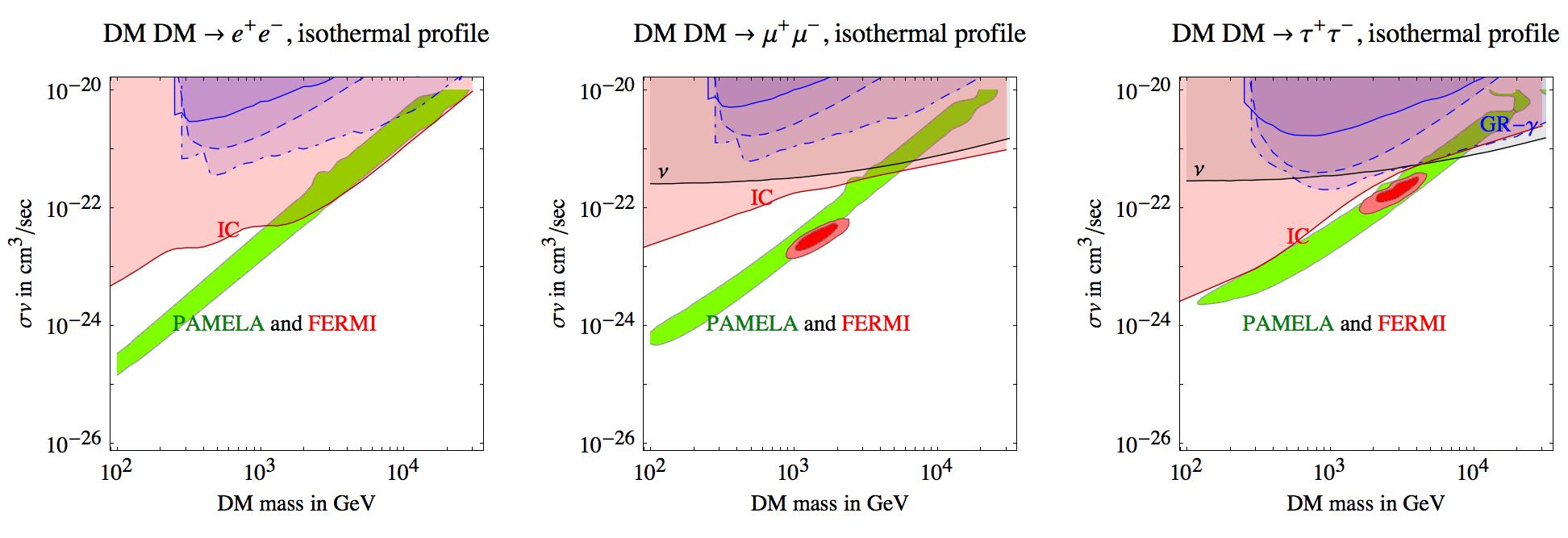}$$
\caption{\em {\bf Direct DM annihilation}.
We compare the region favored by PAMELA (green bands)
and by PAMELA, FERMI and HESS observations (red ellipeses) with
HESS observations of the Galatic Center~\cite{HessGC}
(blue continuous line), Galactic Ridge~\cite{HessGR} (blue dot-dashed),
and spherical dwarfes~\cite{HessSgrDwarf,VERITAS} (blue dashed),
FERMI observations in the `$10^\circ\div20^\circ$' region
and of observations of the Galactic Center at radio-frequencies $\nu=408\,{\rm GHz}$ ~\cite{Davies} 
(dashed red lines) and at $\nu \sim 10^{14}\,{\rm Hz}$ by VLT~\cite{VLT} (upper purple lines, when present, for equipartition and constant magnetic field).  See discussion in the text for remarks regarding the validity of the constraints.
We considered DM annihilations into $e^+ e^-$ (left column), $\mu^+\mu^-$ (middle), $\tau^+\tau^-$ (right),
unity boost and Sommerfeld factors and
the NFW (upper row), Einasto (middle), isothermal (lower) DM density profiles in the Milky Way.
\label{fig:boundse}}
\end{center}
\end{figure}

\begin{figure}[p]
\begin{center}
$$\includegraphics[width=0.99\textwidth]{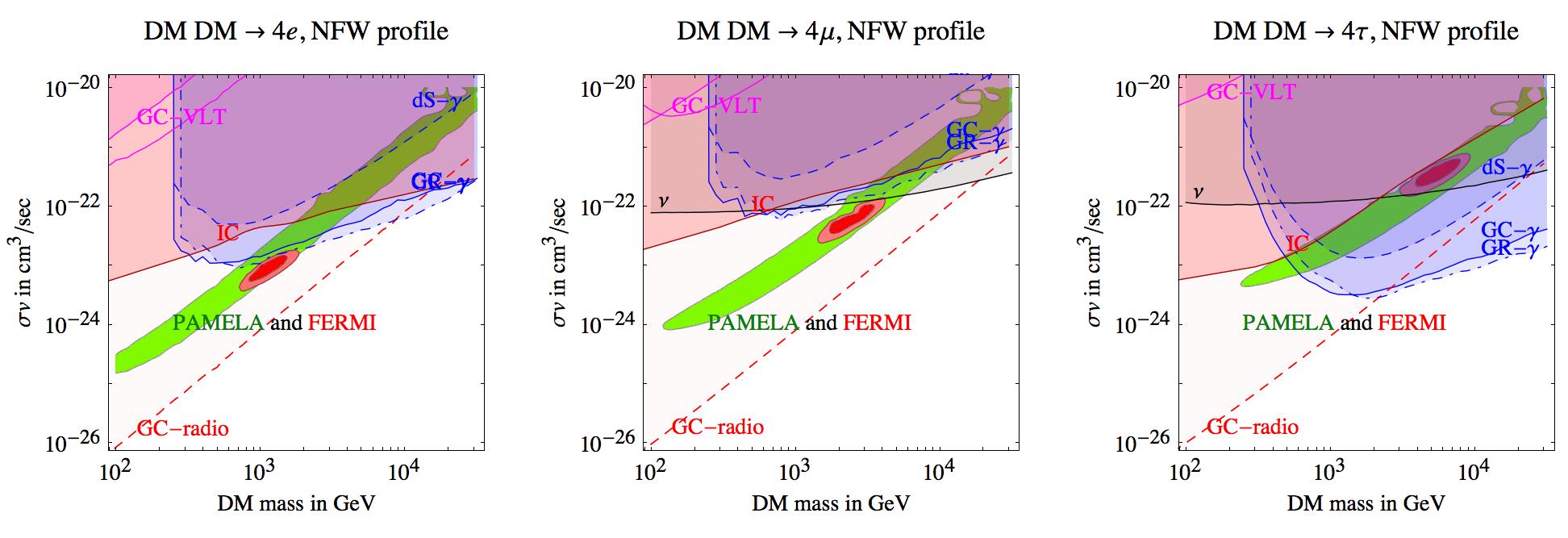}$$
$$\includegraphics[width=0.99\textwidth]{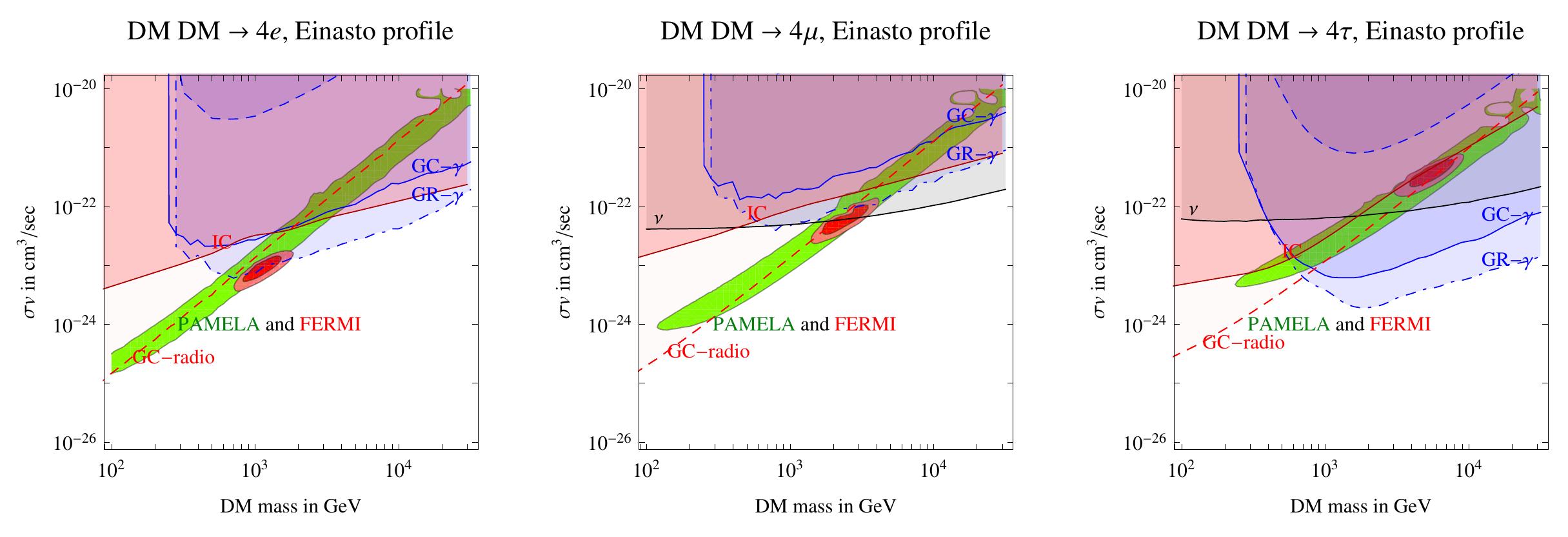}$$
$$\includegraphics[width=0.99\textwidth]{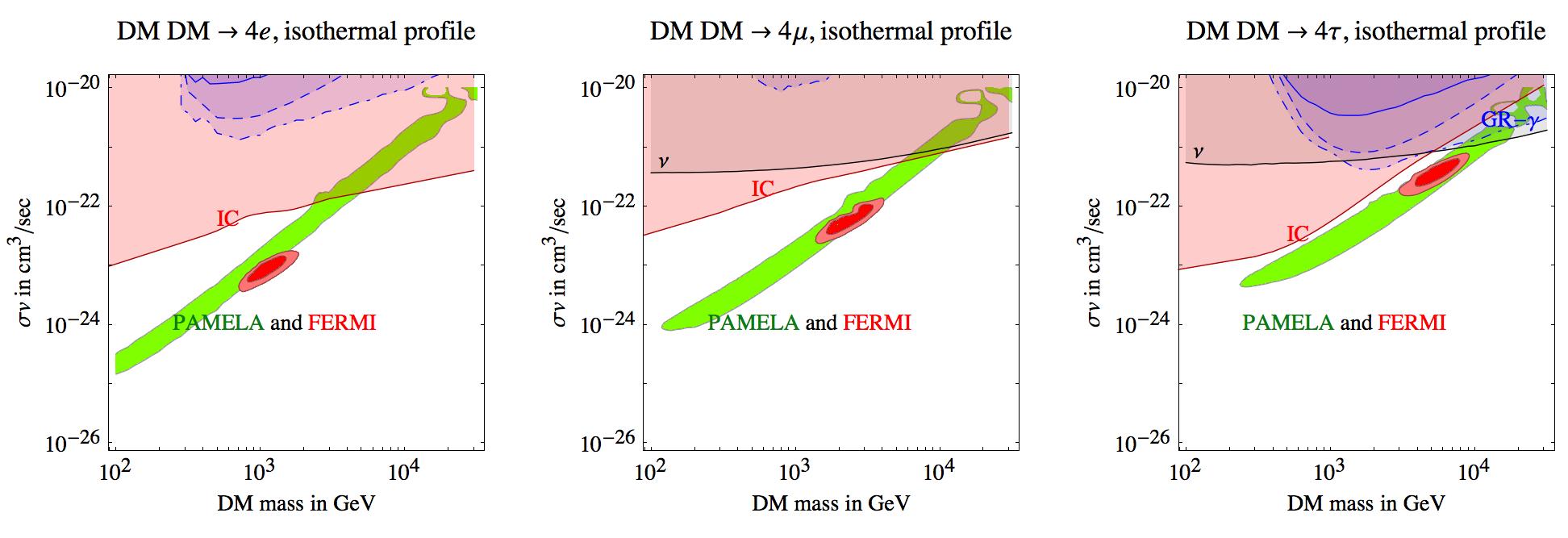}$$
\caption{\em {\bf One-step DM annihilation}.
As in fig.\fig{boundse}, here for DM annihilations into $4e$ (left column), $4\mu$ (middle), $4\tau$ (right),
via a light intermediate new particle.
\label{fig:boundsee}}
\end{center}
\end{figure}

The shaded regions in  the figures are excluded by the various $\gamma$ and $\nu$ constraints:

\begin{enumerate}

\item The blue continuous  curves labeled as `GC-$\gamma$'
shows the bound obtained by imposing that
the HESS observations of {\bf $\gamma$ rays from the Galactic Center} region
(with $0.1^\circ$ angular size) are not exceeded at $3\sigma$ in any data point.
We updated the Galactic Center constraint from~\cite{BCST} in light of the recent preliminary data from HESS~\cite{HessGC},
shown in fig.\fig{ICsamples}d.

\item The blue dot-dashed curves  from~\cite{BCST} labeled as `GR-$\gamma$'
show the corresponding constraint from {\bf ${\gamma}$ observations of the Galactic Ridge} region~\cite{HessGR}.  In view of its larger angular size (a rectangle of size $0.6^\circ \times 1.6^\circ$)
these bounds have a milder dependence on the unknown DM profile.
Again we conservatively imposed that the DM contribution does not exceed any data point at more than $3\sigma$.

\item The blue dashed curves labeled as `dS-$\gamma$'
show the best bound from non-observation of {\bf $\gamma$ from dwarf Spheroidals}.
We proceed as in~\cite{BCST} where the HESS observations of these 
objects were first studied to test
the DM interpretations of  the PAMELA excess, but now adopting the
DM luminosities ${\cal L}$ and the corresponding errors estimated in~\cite{Essig}
and defined such that
\beq \label{eq:L}
\frac{d\Phi_\gamma}{dE_\gamma} = \frac{{\cal L}}{4\pi} \frac{\langle \sigma v\rangle}{2M^2}.
\frac{dN_\gamma}{dE_\gamma}\eeq
For DM annihilations,
in the `NFW' panels we assume the central values of ${\cal L}$;
in the `isoT' panels we assume the $3\sigma$ lower limits on ${\cal L}$
and in the `Einasto' panels we assume the $90\%$ C.L.\ lower limits on ${\cal L}$ as in~\cite{Essig}.
For DM decays, we will in all cases consider the $3\sigma$  lower limits.
We stress that the bounds coming from the Sagittarius Spheroidal dwarf should be taken with caution since its close proximity to the Milky Way causes it to be tidally disrupted which renders the use of conventional DM profiles questionable. 
This is partially reflected by the larger uncertainties in table~\ref{tab:dwarfs}, such that Sagittarius does not provide the dominant constraint.

\begin{table}\small
\begin{center}
\begin{tabular}{cccc}
Spheroidal& DM annihilation  &  DM decay  & bound on the $\gamma$ flux \\
dwarf& log$_{10} (\mathcal{L} \rm{~in~GeV}^2  \rm{cm}^{-5})$ & log$_{10} (\mathcal{L}\rm{~in~GeV} \rm{cm}^{-2})$ &   in $\rm{cm}^{-2}\rm{s}^{-1}$ \\
\hline
Sagittarius  & $19.4 \pm 1.0$ &$18.7 \pm 0.9$ & $\Phi_\gamma(E_\gamma>250 \; \rm{GeV}) < 3.6 \times 10^{-12}$~\cite{HessSgrDwarf}  \\
Draco &$18.6\pm0.4$ & $17.5\pm 0.1$  &$\Phi_\gamma(E_\gamma>200 \; \rm{GeV}) < 2.4 \times 10^{-12}$~\cite{VERITAS}\\
Ursa Minor & $18.8\pm0.8$ & $17.6\pm0.2$  &$\Phi_\gamma(E_\gamma>200 \; \rm{GeV}) < 2.4 \times 10^{-12}$~\cite{VERITAS}\\
Willman 1 & $19.6 \pm 0.6$ & $17.5 \pm 0.5$ &$\Phi_\gamma(E_\gamma>200 \; \rm{GeV}) < 2.4 \times 10^{-12}$~\cite{VERITAS}
\end{tabular}
\caption{\em Central values and $1\sigma$ errors for the
astrophysical factors $\cal L$ defined in eq.\eq{L}
for DM annihilations and decays 
in  Milky Way dwarf galaxies as compiled in~\cite{Essig}.  Note that the bounds coming from Sagittarius should be taken with extreme caution since it is being visibly tidally disrupted by our galaxy, and therefore the DM profile or even ascribing that the dwarf has any DM is potentially suspect.}
\label{tab:dwarfs}
\end{center}
\end{table}

\item {\bf Radio} observation of the inner 4 arc seconds of
the {\bf Galactic Center} constrain synchrotron radiation produced
by $e^\pm$ in the local magnetic fields $B$.
While the uncertainty on $B$ seems not important (as
$e^\pm$ anyhow loose their energy via synchrotron radiation),
the local DM density is very uncertain, so that the resulting bound on DM annihilations
(red dashed curve labeled as `GC-radio', from~\cite{BCST}) is very strong (negligibly weak) if 
a NFW (isothermal) profile is assumed.  
We plot this bound in light colors because it relies on the untested extrapolation that the
growth of the DM density profile as $r\to0$  holds down to $r\sim {\rm pc}$.

\item The black continuous curves labeled as `$\nu$' show the bounds on {\bf neutrinos from the Galactic Center} regions, as observed by SuperKamiokande in regions with angular size
between $3^\circ$ and $30^\circ$: in view of the large observed region such bounds
have a weak dependence on the uncertain DM density profile.
We consider the through-going muon flux and impose that the
contribution due to DM does not exceed the observed rate at $3\sigma$.
In appendix~\ref{nu} we describe how we compute this effect.
As neutrino detection cross sections increase with energy, these constraints
are significant at larger DM masses.

\item Finally, the red continuous curves labeled as `IC' are relative to the preliminary
{\bf FERMI observation in the `$10^\circ\div20^\circ$' region} of $\gamma$.  As discussed  in section~\ref{sec:ics},
in view of the low energy measurement, the DM signal is dominated by ICS $\gamma$ rather than by
Final State Radiation. 
\end{enumerate}

Below we summarize the general lessons learned.

\subsection{DM Annihilations}

Concerning DM annihilations, $2e$, $2\mu$, $2\tau$, $4\tau$ modes are strongly constrained
and are compatible with $\gamma$ bounds only if DM has an effective density profile
that does not significantly grow towards the Galactic Center: either the isothermal profile,
or any profile in models where DM annihilations proceed via extra long-lived particles~\cite{Rothstein:2009pm}.

Only the $4e$ and $4\mu$ modes are (marginally) compatible with the Galactic Ridge HESS observations
also for a NFW or Einasto profile.  Even these cases can already be strongly disfavored by performing
a less conservative global fit of all HESS data points in terms of an astrophysical background plus the DM excess.  This conclusion is changed if hidden sector showering plays a role, as we discuss below. 
Going from $2\ell$ to $4\ell$ (where $\ell=e$ or $\mu$) the FSR $\gamma$ flux is changed in shape~\cite{mpv} and normalization~\cite{mpv,BCST}.  
However, it is difficult to further lower the $\gamma$  constraint below what shown in the $4\mu$ panel.

\begin{figure}[t]
\begin{center}
\includegraphics[width=0.9\textwidth]{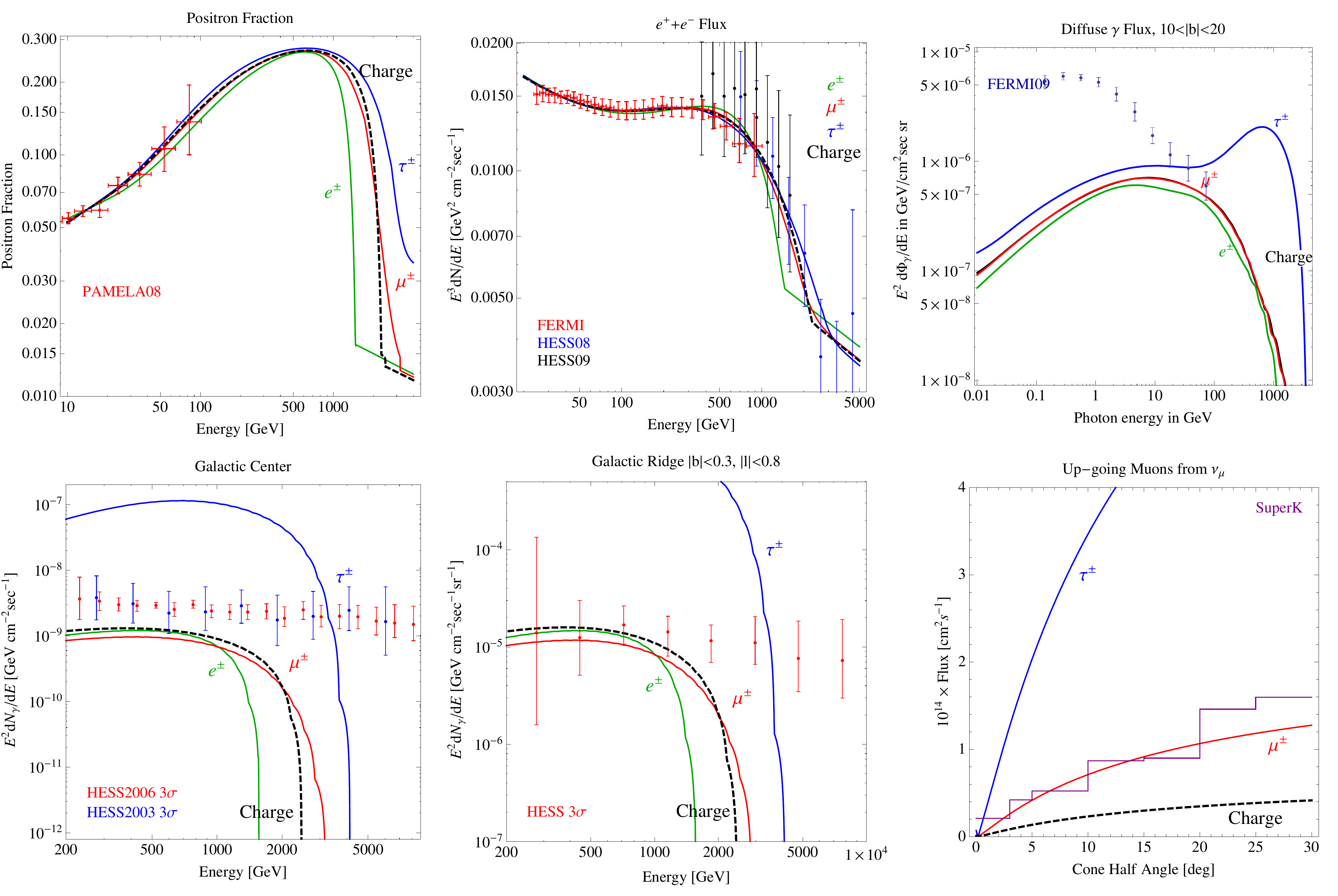}
\caption{\em {\bf Best fits for DM annihilations}.
We assume the MED propagation model and the Einasto profile with $\alpha=0.17$.  We plot the best fit models for DM annihilations into $4e$, $4\mu$, $4\tau$, and final states dictated by coupling through kinetic mixing with the photon ($m_\phi=650\,$ MeV) see figure~\ref{fig:shower}.  All curves include showering with $\alpha_{DM}\sim 0.1$ which increases the goodness of their fit compared to the unshowered spectra.  The plots in the upper row from left to right are for the PAMELA positron fraction, $e^++e^-$ flux recently measured by FERMI and HESS, and the ICS + FSR predictions for these models.  In the bottom row we plot from left to right, the photon predictions for the HESS measurement of the Galactic Center, Galactic Ridge, and the bounds coming from SuperK for those models which create $\nu$'s.
\label{fig:ICsamples}}
\end{center}
\end{figure}

\medskip

The expected neutrino signal is comparable to the present bound on neutrinos from the GC, again favoring a
quasi-flat DM density profile or the $4e$ mode that does not give any neutrinos.
Present SK bounds are limited by statistics but also by the $\nu$ atmospheric background, so that
an improvement would need selecting a sample of higher energy neutrino-induced through-going muons.
This was already attempted in SK, that was able to select a sub-sample of muon-showering events,
namely of muons with $\circa{>}\TeV$ energy in the detector such that muon energy losses are dominated by radiative processes~\cite{SKshower}
(as here discussed in eq.\eq{muEloss}).
Such sub-sample is expected to imply bounds comparable to the bounds considered here.

\begin{figure}[t]
\begin{center}
\includegraphics[width=0.45\textwidth]{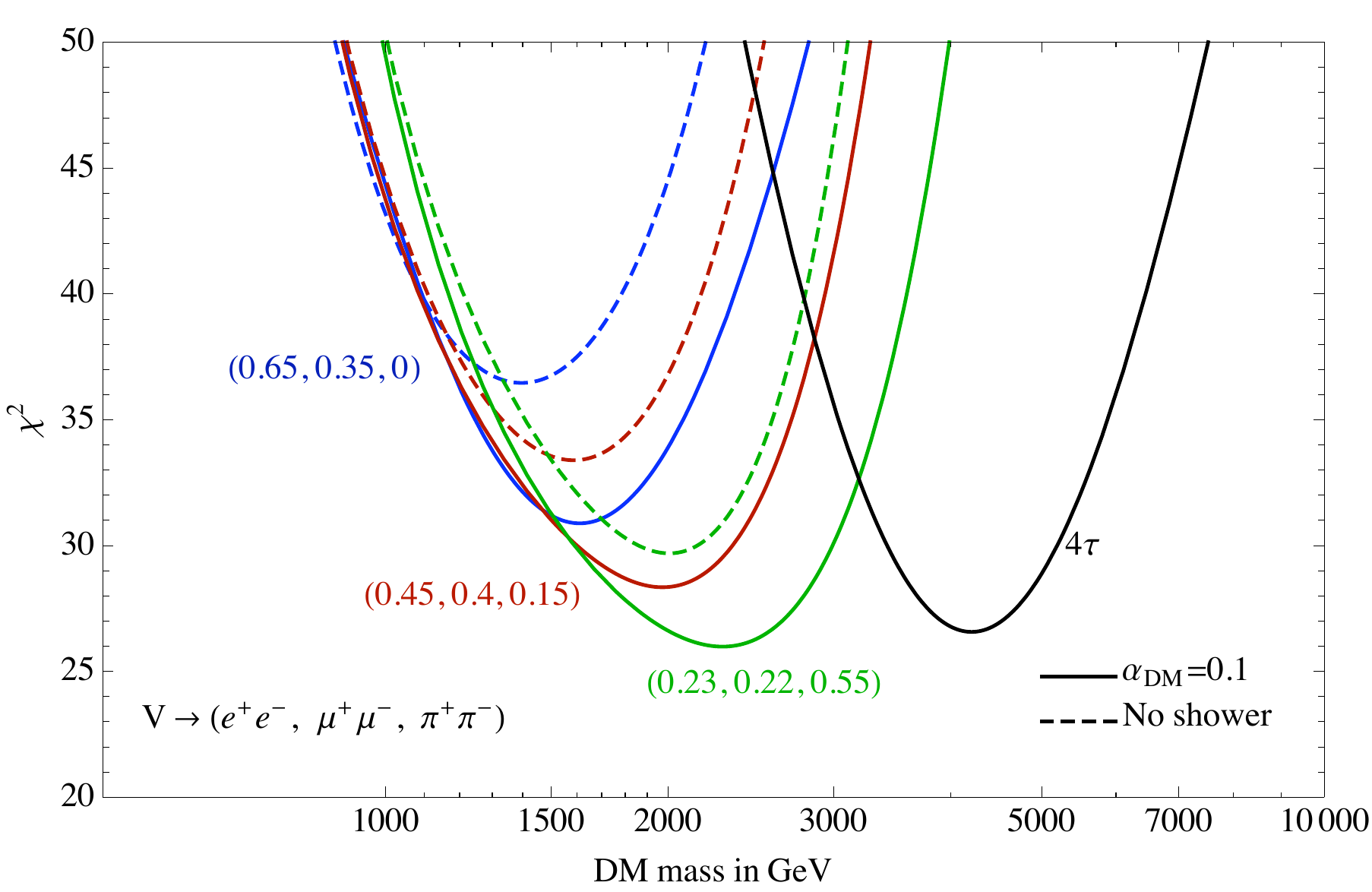}
\includegraphics[width=0.32\textwidth]{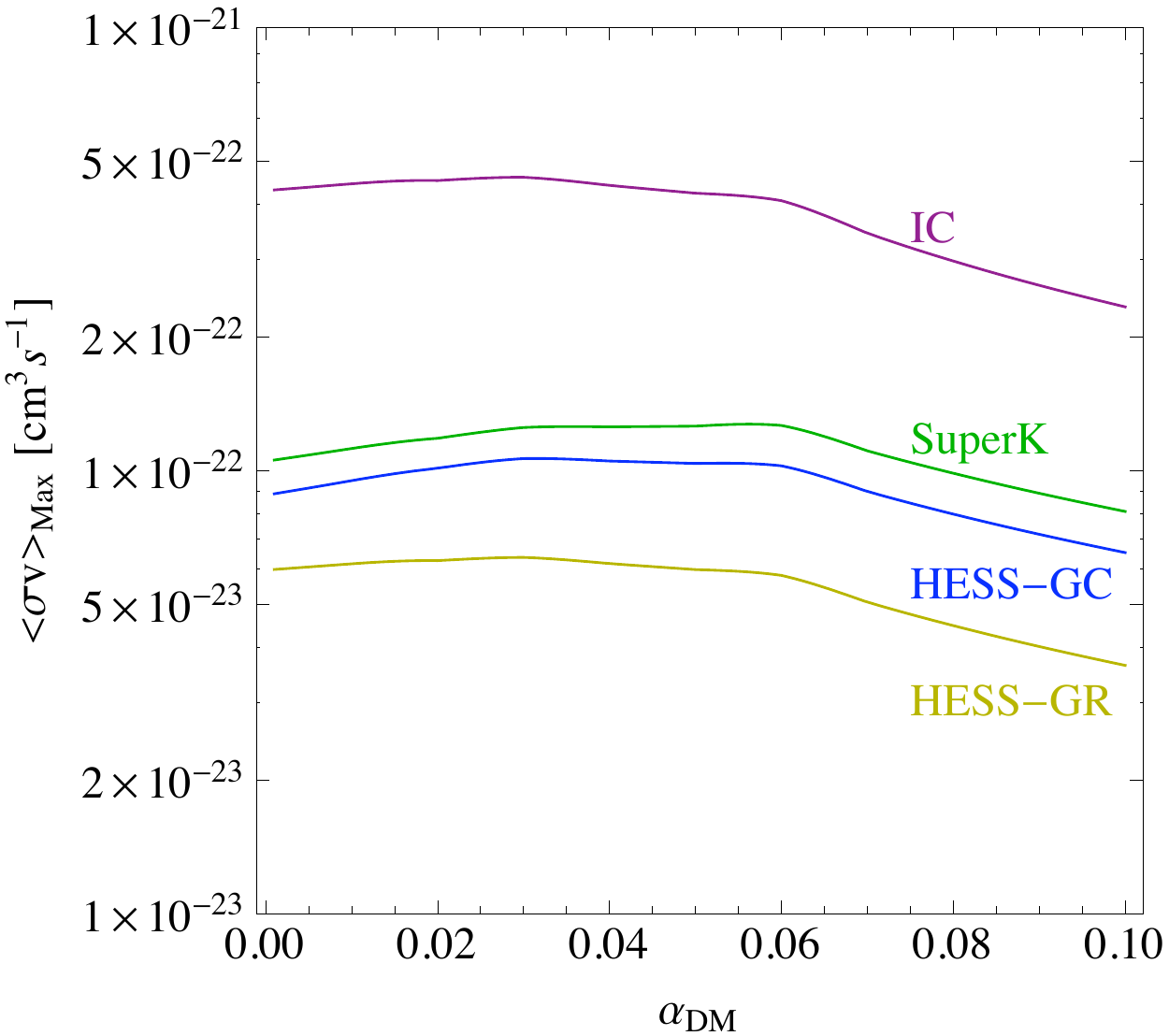}
\caption{\em {\bf Hidden sector shower}.  Left: the $\chi^2$ with and without shower, for various light gauge boson masses of $250$, $450$, and $650 \MeV$ (blue, red, green respectively) that kinetically mix with the photon.  Additionally, a curve for a $4\tau$ final state without shower is shown for comparison of the overall $\chi^2$. 
Right:  Contour lines for various bounds in the $\sigma$ vs $\alpha_{DM}$ plane, for a DM mass of 3 TeV and light gauge boson of mass 650 MeV with an Einasto profile.
 \label{fig:shower}}
\end{center}
\end{figure}

\subsubsection{Hidden Sector Shower}

As discussed in Section 2, an interesting possibility for models that contain a light vector that the DM can annihilate through is a hidden sector shower.  Showering has two primary effects: (i) it softens the spectrum and (ii) it increases the multiplicity.  As we see in Figures~\ref{fig:fit} and \ref{fig:shower}, the hidden sector shower has a nontrivial effect, and improves the $\chi^2$ of our fits.   The hidden sector shower helps reduce the $\chi^2$ by reducing the steepness of the \eflux flux and better fitting the FERMI electrons.  However, as we can see from Fig. \ref{fig:shower}, showering does not improve the $\chi^2$ for all values of $\alpha_{\rm DM}$.  For smaller values of  $\alpha_{\rm DM}$ the spectrum of visible particles is softened without increasing the multiplicity of final state particles greatly.  On the other hand, for larger values of $\alpha_{\rm DM}$, the bounds will strengthen as a function of $\alpha_{\rm DM}$ as the multiplicity becomes too large, generating too many $\gamma$'s.

\medskip

\subsubsection{A Quasi-constant DM Density and Long Lived Intermediate States}
DM annihilations into $2\mu$ and $2\tau$ are still compatible with bounds on the associated $\gamma$ flux if the DM density does not significantly grow towards the Galactic Center.
This possibility is realized in practice by plotting the `isothermal core' DM density profile.  This profile is disfavored by $N$-body simulations and it has no a priori theoretical motivation.  However, in principle there could be some weakening of the more cuspy DM profiles preferred by $N$-body simulations when baryons, which should be important near the center of the galaxy, are included in future simulations.
To explore the effects of a shallower DM density profile
 we have shown in the lower row of Fig.\fig{sample} the best fits for DM annihilations into $2\tau$
assuming the extremal isothermal profile and MED propagation.

For the case of $\ge 4$ SM final states there is one other possibility  to obtain an effective quasi-constant DM density profiles.   
DM can annihilate through a light state that can be long lived on astrophysical scales~\cite{Rothstein:2009pm}.  There is one additional parameter, the lifetime $d$, that effectively acts such that
indirect DM signals must be computed replacing the DM density profile $\rho(r)$  
with its value averaged over a length $d$.
If $d\gg r_\odot$ a constant $\rho$ is effectively obtained.  In our analysis this possibility is effectively
described by the quasi-constant isothermal profile.
For smaller values of $d$  the effective $\rho(r)$ flattens to
a constant only at $r\circa{<} d$ rather than diverging as in the NFW or Einasto profiles; 
the main consequence is relaxing the bounds from $\gamma$ observations from the GC region at $r\circa{<} d$ that we consider.

\begin{figure}[p]
\begin{center}
$$\includegraphics[width=0.99\textwidth]{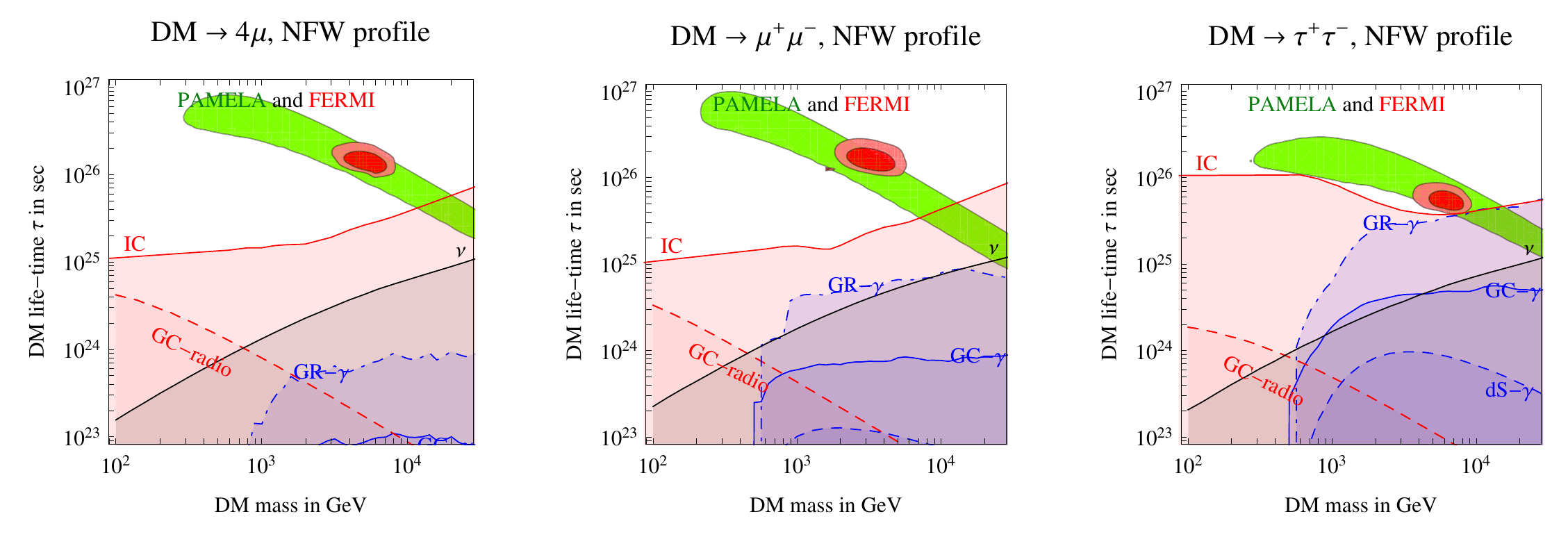}$$
$$\includegraphics[width=0.99\textwidth]{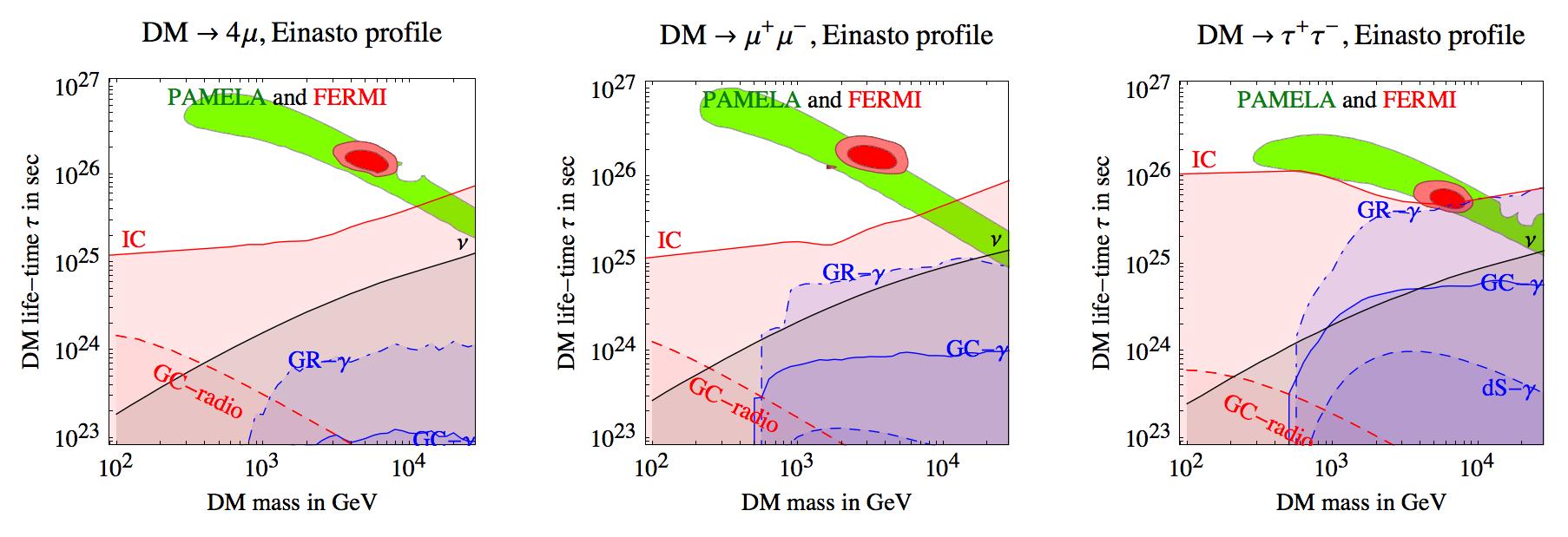}$$
$$\includegraphics[width=0.99\textwidth]{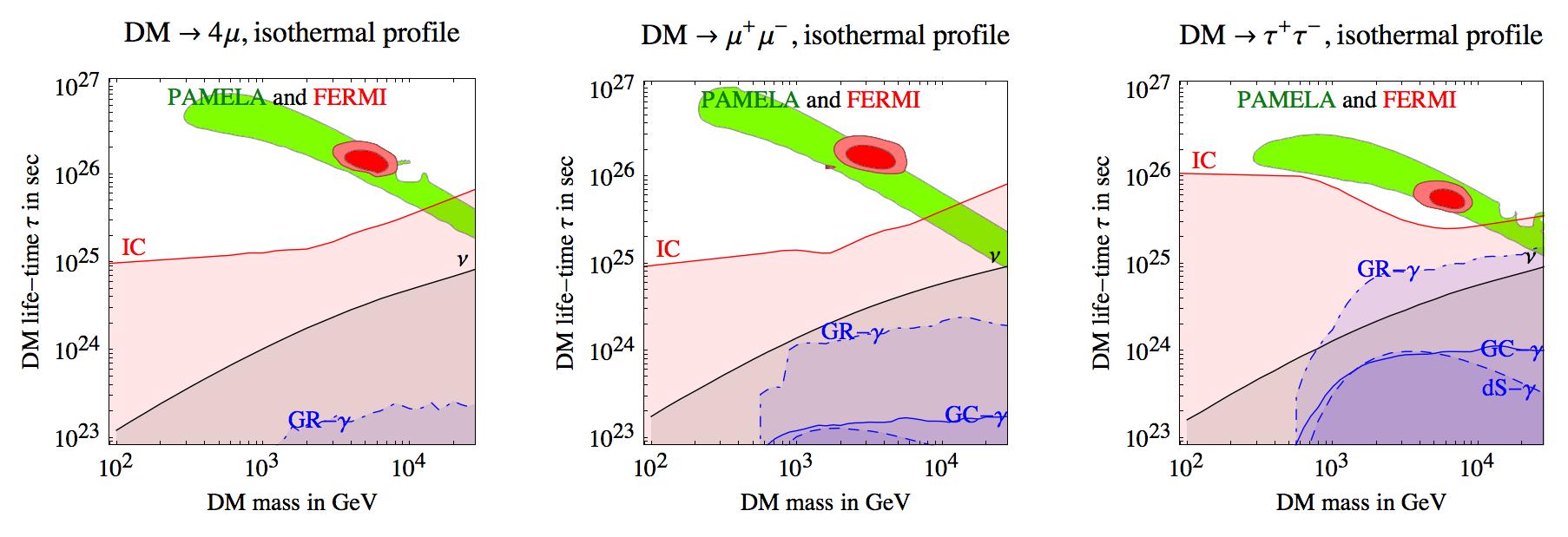}$$
\caption{\em {\bf DM decay}. As in fig.\fig{boundse}, here for DM decaying into  $\mu^+\mu^-$ (middle), $\tau^+\tau^-$ (right),
$4\mu$ (left). We do not consider decay modes into $e^+e^-$, as they do not allow to fit the FERMI data.
\label{fig:boundsDecay}}
\end{center}
\end{figure}

\begin{figure}[t]
\begin{center}
\includegraphics[width=0.9\textwidth]{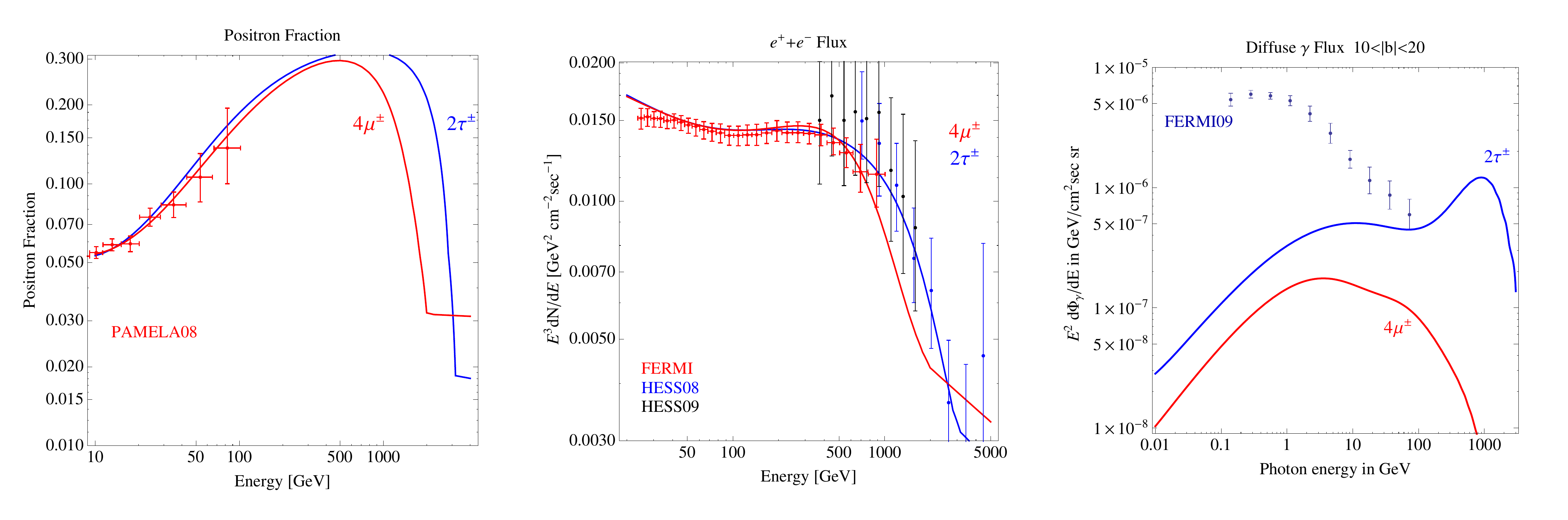}
\caption{\em {\bf Best fits for DM decays}.
As in fig.\fig{sample}, now for DM decays into $4\mu$ (red curve), $2\tau$ (blue curve)
assuming the MED propagation model and the NFW profile.
\label{fig:ICsamplesDecay}}
\end{center}
\end{figure}

\subsection{DM Decays}
The PAMELA and FERMI excesses can be explained if DM decays into leptons with a life-time about $10^9$ times
longer than the age of the universe, which is the typical lifetime of
a TeV-scale particle that decays via a dimension 6 operator suppressed by the GUT scale, $10^{16}\GeV$~\cite{NSS, dimo}.
As discussed in~\cite{NSS} this scenario is compatible with the $\gamma$ bounds even for a NFW scenario,
as the DM decay rate is proportional to the DM density $\rho$, while the annihilation rate is proportional to $\rho^2$
and thereby significantly enhanced close to the Galactic Center.

Here we remark that cosmology offers other constraints on DM annihilations:
BBN~\cite{HisanoBBN}, extra-galactic $\gamma$~\cite{Profumo},
gas heating~\cite{Hooper} and reionization~\cite{Hooper, reionization}. 
These constraints  are significant thanks to the fact that the DM annihilation rate was enhanced in the early universe by the square
of the larger DM density.
The latter constraint is solid, while the first three ones are uncertain, as they depend on the unprecisely
known structure formation history at redshift $z\circa{<}100$.
These constraints are less significant in the alternative interpretation in terms of DM decays.

We here assume that DM decays with 1/2 branching ratio in leptons and in anti-leptons,
although DM present thanks to a baryon-like asymmetry could decay only into anti-leptons.

Fig.\fig{fit}b shows that DM decays can fit the PAMELA, FERMI and HESS $e^\pm$ excesses
as well as DM decays, with some minor differences, mostly due to the fact that 
only in the DM annihilation case 
a sizable amount of lower-energy $e^\pm$ can reach us from the Galactic Center,
giving rise to a smoother $e^\pm$ energy spectrum.
Decay modes of fermionic DM into $W^\pm \mu^\mp$ or $W^\pm \tau^\mp$
provide good fits to $e^\pm$ observations,  but together with a  $\bar{p}$ excess
which is strongly disfavored by  PAMELA $\bar{p}$ observations~\cite{CKRS}.

Fig.\fig{boundsDecay} and \fig{ICsamplesDecay} show that DM interpretations of the PAMELA, FERMI and HESS $e^\pm$ data 
in terms of leptonic DM decays are
compatible with all the constraints we considered, the strongest one being relative to
ICS from the `$10^\circ\div 20^\circ$' region observed by FERMI. 
Other regions (so far observed only by EGRET, with a problematic energy calibration)
offer possibly stronger constraints.   In the near future FERMI will release data in other regions that can be used to constrain this scenario.  As discussed in section~\ref{sec:ics}, one can easily extrapolate the fluxes to other regions using the approximations we have employed.

\subsection{Light Dark Matter and the Electron Component}
DM lighter than about a TeV is {\it firmly excluded in a model-independent way} as an interpretation of the PAMELA excess.
Indeed according to PAMELA, the positron fraction reaches $\sim 15\%$, 
so that a drop of at least $\circa{>} 20\%$ should have been present in the $e^++e^-$ spectrum
at an energy just below the DM mass.
This is not seen in the FERMI data, which have a $\sim1\%$ statistical uncertainty together with
a $\sim 5\%$ systematic uncertainty that cannot fake this drop.
This conclusion was already derived in~\cite{CKRS} in the light of the less precise ATIC data, confirmed at lower energies by FERMI.

Similarly, the lack of an edge in the $e^++e^-$ spectrum implies that
DM that dominantly annihilates or decays into $2e$ is now firmly excluded, and one can constrain
the BR of the subdominant $e^\pm$ primary channel.
Assuming that DM annihilates into leptons with lepton flavor components ${\rm BR}_e,{\rm BR}_\mu,{\rm BR}_\tau$ that sum to unity,
${\rm BR}_e + {\rm BR}_\mu + {\rm BR}_\tau=1$ one can constrain the allowed regions in light of the PAMELA, FERMI and HESS data together with the other constraints.    To clarify the situation we show the constraints on the branching fractions for  three distinct cases:

\begin{enumerate}
\item {\em DM annihilating into $4\ell$  with Einasto profile.}  HESS and SK provide strong constraints
on the associated $\gamma$ and $\nu$ fluxes, which must be taken into account.

\item {\em DM annihilating into $2\ell$ with isothermal DM profile.} Steeper profiles are excluded by HESS and  we therefore restrict ourselves to this profile only.
\item {\em DM decaying into $2\ell$ with NFW profile.}  As shown above, decaying DM evades all bounds and is therefore viable for all 
DM density profiles.
\end{enumerate}

Fig.\fig{triangle} shows the fit for these cases, marginalized
over the DM mass and other parameters.
The FERMI data has been fitted conservatively.  Nonetheless we find that FERMI disfavors DM that annihilates democratically into all $\ell^+ \ell^-$ with 
branching ratios equal to 1/3 (point in the center of the triangle).  This is the situation in many flavor-universal models.
A similar conclusion holds for DM decays into $W^\pm \ell^\mp$.
On the contrary the $4e$ mode is more compatible with data, as it gives a smoother termination of the  $e^\pm$ excess at $E\sim M$.

\begin{figure}[t]
\begin{center}
\includegraphics[width=0.85\textwidth]{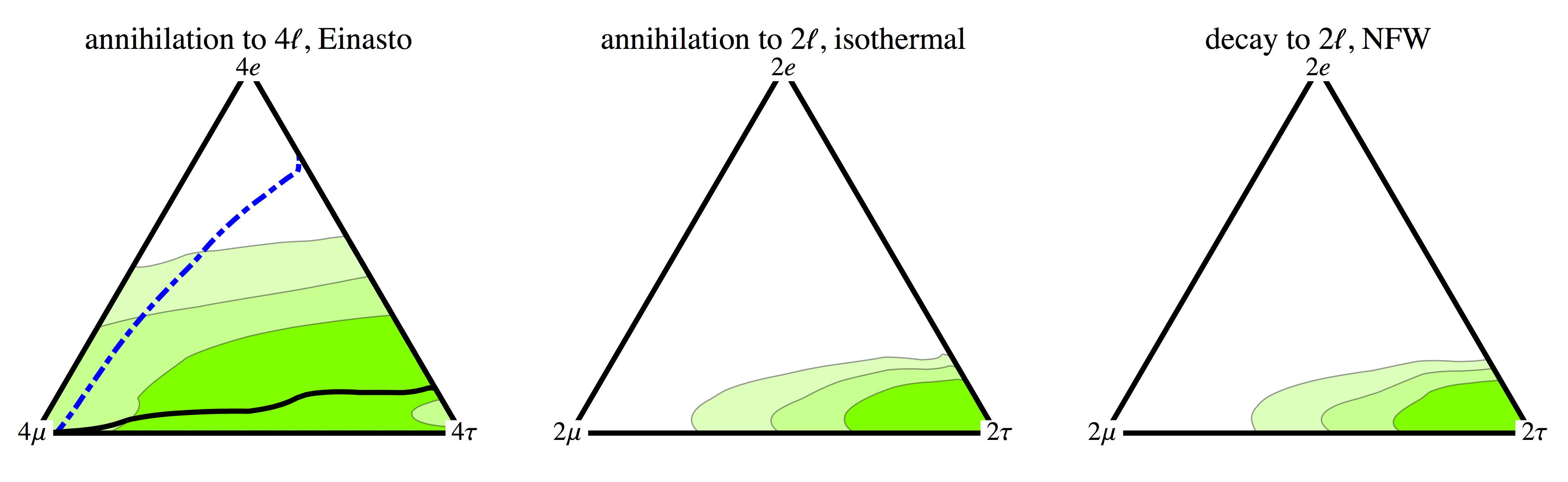}
\caption{\em {\bf Fits for the lepton flavor component} in DM annihilations.
The contour lines correspond to $90\%,99\%,99.9\%$ confidence levels for 2 d.o.f.
In the left panel we consider annihilation into $4\ell$ modes with Einasto profile.  In the center we show  $2\ell$ modes with isothermal profile, and on the right we show decays into $2\ell$ modes with isothermal profile.  Pure flavors are obtained at the corners of the triangles, and the distances ${\rm BR}_\ell$ from the opposite sides (which sum up to unity) indicate the flavor proportion at each point.  
The mostly-$\tau$ region below the dashed (solid thick) curves is excluded by the Galactic Ridge $\gamma$ ($\nu$)   bounds. 
\label{fig:triangle}}
\end{center}
\end{figure}

\section{Conclusions}
We explored how DM can interpret the  excesses in $e^\pm$ cosmic rays measured by FERMI, PAMELA and HESS.   The results of the HESS and FERMI data indicate two specific things for any DM interpretation.
First, the spectrum becomes steeper at around 1 TeV implying the scale of DM needs to be around a TeV.  Second, the FERMI data is smooth from 100 GeV to around a TeV implying that a light DM explanation of PAMELA is {\em inconsistent} with the FERMI data.

We have found that the PAMELA, FERMI and HESS data can be interpreted in terms of admixtures of leptonic final states for DM annihilations or decays. We considered the $e^+e^-$, $\mu^+\mu^-$, $\tau^+\tau^-$ $4e$, $4\mu$, $4\pi$ and $4\tau$ modes.
\begin{itemize}
\item The $e^+e^-$ mode is excluded.
\item Modes involving $\tau$ provide one of the best fits to the $e^\pm$ spectra.
However the  $\tau$ decays into $\pi^0$ imply a large $\gamma$ flux, so that $\tau$ modes
are compatible with bounds from $\gamma$ observations only if
a) DM has a quasi-constant density profile, disfavored by $N$-body simulations; or
b) DM annihilates into a particle with an astro-physically long life-time; or
c) DM decays.
\item
Annihilations into $4\mu$ provide a good fit to the $e^\pm$ spectra
(for $M\approx 3\TeV$) and are marginally compatible with $\gamma$ and $\nu$ bounds for the Einasto profile favored by $N$-body simulations. Again, decays are cleanly compatible with all constraints.

\item The $4e$ mode provides a poorer fit to the FERMI spectrum.
\item Annihilations into $4$ body final states that are dictated by a hidden sector coupling to the SM proportional to charge are marginally consistent with all data.
\item Models that  have a hidden sector shower can provide a better fit, by smoothing the $e^\pm$ spectra.
\end{itemize}
Annihilation modes involving quarks, heavy vectors or the Higgs are disfavored.
A precise observation of the $e^\pm$ excess around its end-point at $\sim 2\TeV$ would allow
to settle this issue and to partially disentangle the allowed leptonic modes.  Additionally for all models that are consistent, the PAMELA positron fraction should continue to increase over the entire range of energy that PAMELA can explore.

We find that, in view of the FERMI observation,
the energy spectrum of photons due to Inverse Compton scattering of $e^\pm$ can now be robustly predicted.
 FERMI should be able to see a $\gamma$ excess if the $e^\pm$ are generated by DM in the DM halo, rather than locally (e.g.\ by a nearby pulsar).  
 This allows one in principle to probe the DM interpretation of the $e^\pm$ excesses and possibly learn about the DM density profile.
To extract further information about the particle physics properties of DM that can explain the $e^\pm$ excesses one would need to observe or more strongly bound the associated neutrino and final state radiation $\gamma$ fluxes.

\paragraph{Acknowledgements} 
We thank Ronaldo Bellazzini, Marco Casolino, Marco Cirelli, Dario Grasso, David Morrissey, Igor Moskalenko, Pierre Salati, David Poland, Martti Raidal, and Neal Weiner for useful discussions.  PM and TV are supported in part by DOE grant DE-FG02-90ER40542. MP is supported in part by NSF grant PH0503584. Research funds of AS have been in toto expropriated by his Department in financial difficulties.

\medskip 

\noindent{\bf Note added:} While this paper was in preparation a related work appeared~\cite{Bergstrom:2009fa}.

\begin{figure}
\begin{center}
\includegraphics[width=0.45\textwidth]{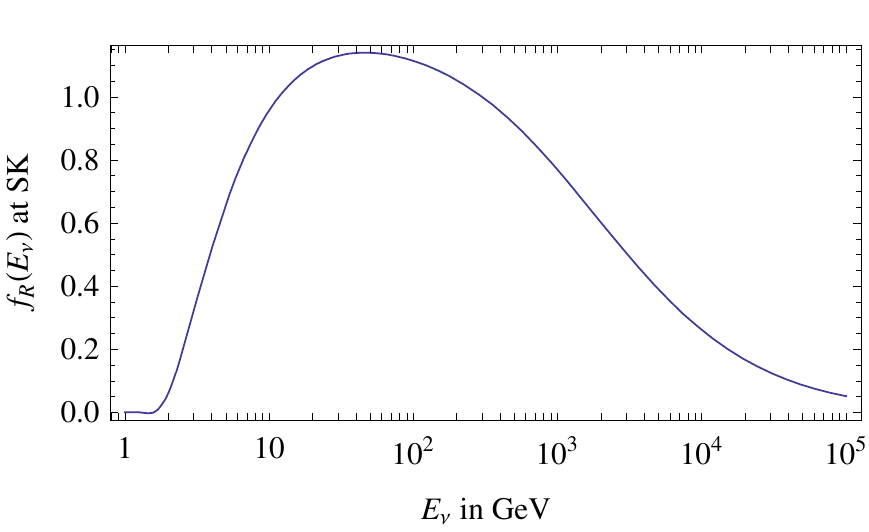}
\caption{\em Function $f_R(E_\nu)$ relevant for the neutrino signals.
\label{fig:fnu}}
\end{center}
\end{figure}

\appendix\section{Neutrinos from the Galactic Center}\label{nu}
SK observes the flux of through going (TG) muons, generated by $\nu_\mu$ scatterings in the matter below the SK detector
down to a depth given by the muon range.
We recall that muon energy losses and range $R$ in matter can be approximated as
\beq \label{eq:muEloss}\frac{dE_\mu}{dx} = \alpha_\mu + \beta_\mu E_\mu,\qquad
R (E_\mu)= \frac{1}{\beta}\ln (1+\frac{\beta_\mu}{\alpha_\mu}E_\mu)\eeq
where  $\alpha_\mu = 2\cdot 10^{-3}\, {\rm GeV g}^{-1}\,{\rm cm}^2$ and $\beta_\mu =3\cdot 10^{-6} {\rm g}^{-1}\,{\rm cm^2}$
becomes dominant at $E_\mu \circa{>}\TeV$.

The $\nu_\mu N \to \mu N'$ cross section is
\beq\frac{d\sigma}{dE_\mu} = \frac{2G_{\rm F}^2 m_N}{\pi}\left[ p_1 + p_2 \left(\frac{E_\mu}{E_\nu}\right)^2\right]\eeq
where $p_{1,2}$ describe the parton content of the nucleon:
$p_1 \approx 0.2$ and $p_2 \approx 0.05$ for $\nu$, and the reverse for $\bar\nu$, having assumed an average nucleon $N$
with equal number of $n$ and $p$ and $E_\nu \ll M_W^2/m_N\sim 10\TeV$.
Then, the total number of TG muons produced by a generic energy distribution $dN_\nu/dE_\nu$ of neutrinos is:
\beq  N_\mu^{\rm TG} = \int_0^\infty dE_\nu \frac{dN_\nu}{dE_\nu} f(E_\nu)\eeq
where the dimensionless function,  $f$,  is given by
\beq
f(E_\nu)=
\lambda \int_{E_{\rm tresh}}^{E_\nu} dE_\mu \frac{d\sigma}{dE_\mu} [R(E_\mu)-R( E_{\rm tresh})]    \eeq
and can be written as,
\beq f(E_\nu) = {E_\nu^2}\frac{G_{\rm F}^2 (2 p_1+p_2)}{2\pi\alpha_\mu} f_R(E_\nu) .\eeq
Here $f_R=1$ as long as $E_\mu \ll \TeV$ such that one can approximate $\beta_\mu=0$.  We plot the function $f_R$ in Fig.\fig{fnu}.
It vanishes at negligibly small neutrino energies due to the (negligibly small) threshold $E_{\rm tresh}=1.6\GeV$ in SK.

We therefore find,
\beq
N_\mu^{\rm TG}=  \frac{\langle\sigma v\rangle}{2}\frac{r_\odot}{4\pi} \frac{\rho_\odot^2}{M^2} J S_F\times
\frac{M^2 G_{\rm F}^2 (2p_1+p_2)}{2\pi \alpha_\mu},
\qquad S_F = \int_0^1 dx~x^2 \frac{dN_\nu}{dx} f_R.\eeq
Notice that the dependence on $M$ cancels out, up to the residual dependence implicit in the factor $f_R$.
Up to this factor our formul\ae{} confirm the ones in~\cite{HisanoNu}.

\bigskip

\section{Details of the Fitting Procedure}\label{fit}
We smoothly vary between the NFW, Einasto ($\alpha=0.17)$ and isothermal DM density profiles,
defined as in~\cite{BCST}.
In particular, we keep fixed the local DM density $\rho_\odot = 0.3\GeV/\cm^3$, which is uncertain
by about a factor of 2.
Since the indirect signals are proportional to $\langle \sigma v\rangle \rho^2$ or to $\rho/\tau$
for DM annihilations and decays respectively, different values of $\rho_\odot$ can be studied by rescaling
$\langle \sigma v\rangle$ or $\tau$.

We approximate the diffusion zone with a cylinder of half-thickness $L$ centered on the galactic plane
and adjust the $e^\pm,\bar p$ diffusion parameters as function of $L$ in order to reproduce cosmic ray data.
We smoothly vary between the favored case $L=4$ kpc (MED configuration) and the extremal cases
$L=1$ kpc (MIN configuration) and $L=15$ kpc (MAX configuration).
All details are as in~\cite{CKRS}.

Concerning the astrophysical backgrounds, we fit in two ways.  First in the method similar to~\cite{CKRS}, which employs the background estimate from~\cite{Strong}, allowing for an uncertainty in the overall normalization and the spectral index of each component.
The new FERMI data allow us to now enlarge the uncertainties, as the fit now relies more on data rather than the theoretical computations.  Second we use an almost completely data driven approach to backgrounds as described in~\cite{mpv}.  We find that for all fits these two methods give similar results and we consider our background estimates to be as robust as possible without the ability to determine a better estimate of backgrounds independently.

We checked that including the latest 2009 HESS $e^++e^-$~\cite{HESSepm}  data negligibly affects our fits,
as the latest HESS data extend to lower energies where FERMI provides more precise observations.
Similarly, the preliminary PAMELA measurement of the un-normalized $e^-$ flux~\cite{waseda} similarly has a negligible impact
on the global fits, in view of the more precise FERMI measurements.

\bigskip\newpage

\footnotesize

\begin{multicols}{2}

\end{multicols}

\end{document}